\documentclass[aip,jcp,preprint]{revtex4-1}%
\usepackage{amsmath,amsfonts,amsthm}
\usepackage{placeins}
\usepackage{tabularx,multirow}
\usepackage[T1]{fontenc}
\usepackage{graphicx}
\usepackage[english]{babel}
\usepackage{amsmath}
\usepackage{amsfonts}
\usepackage{amssymb}
\usepackage{xcolor}
\usepackage{natbib}
\usepackage{dcolumn}%
\providecommand{\U}[1]{\protect\rule{.1in}{.1in}}
\graphicspath{{./Figures/}{D:/Papers/Konstantin/EIE_extended_space/Figures/}
{C:/Users/Jiri/Dropbox/Papers/Chemistry_papers/2016/EIEs_stochastic/Revision/Figures/}}
\makeatletter
\providecommand{\tabularnewline}{\\}

\@ifundefined{textcolor}{}
{
\definecolor{BLACK}{gray}{0}
\definecolor{WHITE}{gray}{1}
\definecolor{RED}{rgb}{1,0,0}
\definecolor{GREEN}{rgb}{0,1,0}
\definecolor{BLUE}{rgb}{0,0,1}
\definecolor{CYAN}{cmyk}{1,0,0,0}
\definecolor{MAGENTA}{cmyk}{0,1,0,0}
\definecolor{YELLOW}{cmyk}{0,0,1,0}
}
\makeatother
\begin{document}
\title{Accelerating equilibrium isotope effect calculations: I. Stochastic
thermodynamic integration with respect to mass}
\author{Konstantin Karandashev}
\email{konstantin.karandashev@epfl.ch}
\author{Ji\v{r}\'{\i} Van\'{\i}\v{c}ek}
\email{jiri.vanicek@epfl.ch}
\affiliation{Laboratory of Theoretical Physical Chemistry, Institut des Sciences et
Ing\'{e}nierie Chimiques, \'{E}cole Polytechnique F\'{e}d\'{e}rale de Lausanne
(EPFL), CH-1015, Lausanne, Switzerland}
\date{March 9, 2017}

\begin{abstract}
Accurate path integral Monte Carlo or molecular dynamics calculations of
isotope effects have until recently been expensive because of the necessity to
reduce three types of errors present in such calculations: statistical errors
due to sampling, path integral discretization errors, and thermodynamic
integration errors. While the statistical errors can be reduced with virial
estimators and path integral discretization errors with high-order
factorization of the Boltzmann operator, here we propose a method for
accelerating isotope effect calculations by eliminating the integration error.
We show that the integration error can be removed entirely by
changing particle masses stochastically during the calculation and by using a
piecewise linear umbrella biasing potential. Moreover, we demonstrate
numerically that this approach does not increase the statistical error. The
resulting acceleration of isotope effect calculations is demonstrated on a
model harmonic system and on deuterated species of methane.

\end{abstract}
\maketitle

\section{Introduction}

The equilibrium (or thermodynamic) isotope effect\cite{Wolfsberg_Rebelo:2010}
is defined as the effect of isotopic substitution on the equilibrium constant
of a chemical reaction. More precisely, the equilibrium isotope effect is the
ratio of equilibrium constants,%
\begin{equation}
\mathrm{EIE}:=\frac{K^{(B)}}{K^{(A)}},
\end{equation}
where $A$ and $B$ are two isotopologues of the reactant. Since an
equilibrium constant can be evaluated as the ratio of the product and reactant
partition functions ($K=Q_{\text{prod}}/Q_{\text{react}}$), every equilibrium
isotope effect can be written as a product of several \textquotedblleft
elementary\textquotedblright\ isotope effects (IEs),
\begin{equation}
\mathrm{IE}=\frac{Q^{(B)}}{Q^{(A)}}, \label{eq:IE_definition}%
\end{equation}
given by the ratio of partition functions corresponding to different
isotopologues (of either the reactant or product).

This quantity is closely related to the important notion of \emph{isotope
fractionation},\cite{Urey:1947,Wolfsberg_Rebelo:2010,McKenzie_Ramesh:2015}
which describes the distribution of isotopes in different substances or
different phases and can be expressed in terms of such elementary isotope
effects~(\ref{eq:IE_definition}) if kinetic factors can be neglected. Below,
we will therefore focus on finding these elementary ratios of partition
functions and call them ``isotope effects'' for short.

The isotope effect is extremely useful in uncovering the influence of nuclear
quantum effects on molecular
properties,\cite{Janak_Parkin:2003,Wolfsberg_Rebelo:2010,McKenzie_Ramesh:2015}
hence many approaches have been developed to calculate it. The
simplest and most common approach, usually referred to as the
\textquotedblleft harmonic approximation\textquotedblright\ or
\textquotedblleft Urey model\textquotedblright, assumes (i) separability of
rotations and vibrations, (ii) rigid rotor approximation for the rotations,
and (iii) harmonic oscillator approximation for the
vibrations.\cite{Urey:1947,Wolfsberg_Rebelo:2010,Webb_Miller:2014} Although
there exist various corrections that incorporate the leading effects of
rovibrational coupling, nonrigidity of the rotor, or anharmonicity of the
vibrations,\cite{Richet_Javoy:1977,Barone:2004,Liu_Liu:2010} this perturbative
approach is not always sufficient; indeed, there are examples of systems in
which these corrections can even yield worse results than the Urey
model.\cite{Webb_Miller:2014}

\begin{figure}
[tbp]\centering\includegraphics[width=\textwidth]{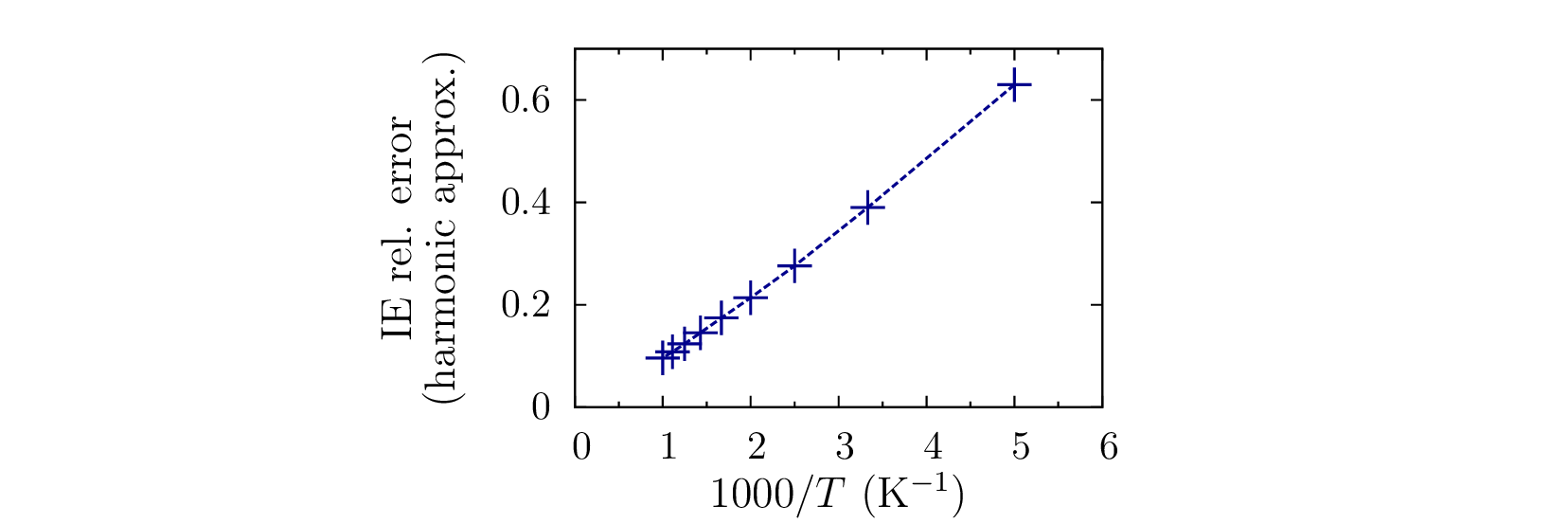}
\caption{Temperature dependence of the relative error of the $ \mathrm{CD_{4}/CH_{4}}$ isotope effect (IE) obtained with
the harmonic approximation. The result of stochastic thermodynamic integration (STI, see Subsec.~\ref{subsec:methane_deuterization}) serves
as a reference; the relative error is defined as IE(harmonic)/IE(STI) - 1.
} \label{fig:ch4_PI_vs_harm}
\end{figure}

We therefore employ a more rigorous method that avoids these
approximations altogether and treats the potential energy surface, rotations,
and rovibrational coupling exactly. To show the benefit of this rigorous
approach, in Figure~\ref{fig:ch4_PI_vs_harm} we plot the relative error of
$\mathrm{CD}_{4}/\mathrm{CH}_{4}$ IE calculated with the harmonic
approximation. In this example, the harmonic approximation works rather well
at higher temperatures, where the IE is small, but its error reaches as much
as $60\%$ at the low temperature of $200$\thinspace K, where the IE becomes
very large.

The potentially large errors of the harmonic approximation are
eliminated in the Feynman path integral formalism,\cite{Feynman_Hibbs:1965,
Chandler_Wolynes:1981, Ceperley:1995} in which the quantum partition function
is transformed to a classical partition function of the so-called
ring polymer; it is then possible to compute the isotope effect
via the thermodynamic
integration\cite{Kirkwood:1935,Frenkel_Smit:2002,Tuckerman:2010} with respect
to mass,\cite{Vanicek_Aoiz:2005,Vanicek_Miller:2007,
Zimmermann_Vanicek:2009,Zimmermann_Vanicek:2010,Perez_Lilienfeld:2011} which
treats the isotope masses as continuous variables and allows using standard
path integral molecular dynamics or Monte Carlo techniques. The main drawback
of this approach is that the \textquotedblleft mass integral\textquotedblright%
\ is evaluated by discretizing the mass, which introduces an
integration error. Although several elegant tricks reduce this integration
error significantly,\cite{Ceriotti_Markland:2013,Marsalek_Tuckerman:2014} it
can never be removed completely if the integral is evaluated deterministically.

Here we propose a way to bypass this issue by augmenting the configuration
space of the Monte Carlo simulation with an extra dimension $\lambda$
corresponding to the mass and including this dimension in the stochastic
integration. The main idea is quite similar to that employed in the more
general $\lambda$-dynamics
method\cite{Liu_Berne:1993,Kong_Brooks:1996,Guo_Kong:1998,Bitetti-Putzer_Karplus:2003,Perez_Lilienfeld:2011}
used in molecular dynamics or Monte Carlo simulations, but we
introduce a Monte Carlo procedure which is applicable for the specific case of
the change of mass and enables a faster exploration of the $\lambda$
dimension. We find that the proposed stochastic approach reduces
the integration error of the IE drastically without increasing the
statistical error. Remarkably, we also show that the integration error can be
reduced to zero exactly by using a piecewise linear umbrella biasing
potential; the only remaining error of the calculated IE is due to statistical
factors.

To assess the numerical performance of the proposed methodology,
we apply it to the isotope effects in an
eight-dimensional harmonic model and in a full-dimensional $\mathrm{CH}_{4}$
molecule. Methane was chosen because the $\mathrm{CH_{4}+D_{2}}$
exchange is an important benchmark reaction for studying catalysis of hydrogen
exchange over metals\cite{Osawa_Lee:2010} and metal
oxides,\cite{Hargreaves_Taylor:2002} and because the polydeuterated species
$\mathrm{CH_{4-x}D_{x}}$ are formed in abundance during the catalyzed reaction.

\section{Theory}

\label{sec:Theory}

\subsection{Path integral representation of the partition function}

Let us consider a molecular system consisting of $N$ atoms with masses $m_{i}$
($i=1,\ldots,N$) moving in $D$ spatial dimensions (typically, $D=3$, of
course). To apply the path integral formalism to the isotope
effect~(\ref{eq:IE_definition}), one first needs a path integral
representation of the partition function $Q:=\operatorname{Tr}\exp(-\beta
\hat{H})$.

This representation is obtained by factoring the Boltzmann
operator $\exp(-\beta\hat{H})=[\exp(-\beta\hat{H}/P)]^{P}$ into $P$ so-called
imaginary time slices, inserting a coordinate resolution of identity between
each two adjacent factors, using a high-temperature approximation for each
factor $\exp(-\beta\hat{H}/P)$, and taking a limit $P\rightarrow\infty$, in
which the high-temperature approximation becomes exact. The
well-known\cite{Feynman_Hibbs:1965, Chandler_Wolynes:1981} final result%
\begin{equation}
Q=\lim_{P\rightarrow\infty}Q_{P}%
\end{equation}
expresses the partition function as the $P\rightarrow\infty$ limit of the
discretized path integral representation
\begin{equation}
Q_{P}=\int d\mathbf{r}\rho(\mathbf{r})\text{,} \label{eq:Qr_PI}%
\end{equation}
where $\mathbf{r}$ is a vector containing all $PND$ coordinates
of all atoms in all slices of the extended configuration space; more
precisely, $\mathbf{r}:=\left(  \mathbf{r}^{(1)},\ldots,\mathbf{r}%
^{(P)}\right)  $, where $\mathbf{r}^{\left(  s\right)  }$, $s=1,\ldots,P$, is
a vector containing all $ND$ coordinates of all atoms in slice $s$. In
general, a subscript $P$ on a quantity $A$ will denote a discretized path
integral representation of $A$ using $P$ imaginary time slices. The
statistical weight $\rho(\mathbf{r})$ of each path integral
configuration is given by%
\begin{equation}
\rho(\mathbf{r})=C\exp\left[  -\beta\Phi(\mathbf{r})\right]  \text{,}
\label{eq:rho}%
\end{equation}
with the prefactor%
\begin{equation}
C=\left(  \frac{P}{2\beta\hbar^{2}\pi}\right)  ^{NDP/2}\left(  \prod_{i=1}%
^{N}m_{i}\right)  ^{DP/2},
\end{equation}
and with an effective potential energy of
the classical ring polymer given by
\begin{equation}
\Phi(\mathbf{r})=\frac{P}{2\beta^{2}\hbar^{2}}\sum_{i=1}^{N}m_{i}\sum
_{s=1}^{P}|\mathrm{r}_{i}^{(s)}-\mathrm{r}_{i}^{(s-1)}|^{2}+\frac{1}{P}%
\sum_{s=1}^{P}V(\mathbf{r}^{(s)})\text{,} \label{eq:Phi}%
\end{equation}
where $\mathrm{r}_{i}^{(s)}$ denotes the component of
$\mathbf{r}^{(s)}$ corresponding to atom $i$ (i.e. $\mathrm{r}%
_{i}^{(s)}$ is a vector containing the $D$ coordinates of the $i$th atom),
and $V$ is the potential energy of the original system. Since the
factorization of the Boltzmann operator is an example of the Lie-Trotter
factorization, the number $P$ is also referred to as the Trotter number.
Because the path employed to represent the partition function is a closed
path, we define $\mathbf{r}^{(0)}:=\mathbf{r}^{(P)}$; this convention was
already used in Eq.~(\ref{eq:Phi}) for $s=1$.

Note that $Q_{P}$ is a classical partition function of the ring
polymer, i.e., a system in the extended configuration space with $NDP$
classical degrees of freedom and defined by the effective potential
$\Phi(\mathbf{r})$. For $P=1$ the path integral expression
(\ref{eq:Qr_PI}) for the quantum partition function reduces to the classical one.

\subsection{Thermodynamic integration with respect to mass}

\label{subsec:TI}

Our ultimate goal is evaluating the isotope effect~(\ref{eq:IE_definition}),
i.e., a ratio of partition functions. Although it is possible to evaluate
partition functions $Q_{P}^{(A)}$ and $Q_{P}^{(B)}$ themselves with a Monte
Carlo procedure,\cite{Lynch_Truhlar:2005} it is more convenient to calculate
the ratio $Q_{P}^{(B)}/Q_{P}^{(A)}$ directly. We now review the most common of
such direct approaches, based on thermodynamic integration\cite{Kirkwood:1935}
with respect to mass.\cite{Vanicek_Aoiz:2005}

In this method, it is assumed that the isotope change is continuous and
parametrized by a dimensionless parameter $\lambda\in\lbrack0,1]$, where
$\lambda=0$ corresponds to isotopologue $A$ and $\lambda=1$ to
isotopologue $B$. This allows, e.g., the description of the
isotope effect when several atoms in a molecule are replaced by their isotopes
simultaneously. Therefore we define, for each atom $i$, a
continuous function $m_{i}(\lambda)$ of $\lambda$ such that
\begin{align}
m_{i}(0)  &  =m_{i}^{(A)},\\
m_{i}(1)  &  =m_{i}^{(B)}.
\end{align}
The simplest possible choice for the interpolating function is the linear
interpolation
\begin{equation}
m_{i}(\lambda)=\left(  1-\lambda\right)  m_{i}^{(A)}+\lambda m_{i}^{(B)},
\label{eq:m_linear_interpol}%
\end{equation}
used in
Refs.~\onlinecite{Vanicek_Aoiz:2005,Vanicek_Miller:2007,Zimmermann_Vanicek:2009},
but Ceriotti and Markland\cite{Ceriotti_Markland:2013} showed that a faster
convergence, especially in the deep quantum regime, is often achieved by
interpolating the inverse square roots of the masses,
\begin{equation}
\frac{1}{\sqrt{m_{i}(\lambda)}}=\left(  1-\lambda\right)  \frac{1}{\sqrt
{m_{i}^{(A)}}}+\lambda\frac{1}{\sqrt{m_{i}^{(B)}}}, \label{eq:m_root_interpol}%
\end{equation}
which is therefore the interpolation used in the numerical
examples below, unless explicitly mentioned otherwise.

Letting $Q(\lambda)$ denote the partition function of a
fictitious system with interpolated masses $m_{i}(\lambda)$, we
can express the isotope effect (\ref{eq:IE_definition}) as
\begin{align}
\frac{Q^{(B)}}{Q^{(A)}}  &  =\exp\left[  \int_{0}^{1}\frac{d\ln Q(\lambda)
}{d\lambda}d\lambda\right] \label{eq:introducing_thermodynamic_integration}\\
&  =\exp\left[  -\beta\int_{0}^{1}\frac{dF(\lambda)}{d\lambda}d\lambda\right]
,
\end{align}
where $F(\lambda)$ is the free energy corresponding to the
isotope change and the integral in the exponent motivated the name
\textquotedblleft thermodynamic integration.\textquotedblright\ While it is
difficult to evaluate either $Q_{P}^{(A)}$ or $Q_{P}^{(B)}$ with a path
integral Monte Carlo method, the logarithmic derivative $d\ln
Q_{P}(\lambda)/d\lambda=\left[  dQ_{P}(\lambda)/d\lambda\right]
/Q_{P}(\lambda)=-\beta dF_{P}(\lambda)/d\lambda$ is a normalized quantity,
i.e., a thermodynamic average proportional to the free energy derivative with
respect to $\lambda$, and therefore can be computed easily with the Metropolis
algorithm with sampling weight $\rho^{(\lambda)}(\mathbf{r})$
corresponding to the fictitious system with masses $m_{i}%
(\lambda)$:
\[
dF_{P}(\lambda)/d\lambda=\left\langle \left[  dF(\lambda)/d\lambda\right]
_{\text{est}}\right\rangle ^{(\lambda)}.
\]
Here we have introduced general notation
\[
\left\langle A_{\text{est}}\right\rangle ^{(\lambda)}:=\frac{\int
d\mathbf{r}A_{\text{est}}(\mathbf{r})\rho^{(\lambda)}(\mathbf{r})}{\int
d\mathbf{r}\rho^{( \lambda) }(\mathbf{r})}%
\]
for a thermodynamic path integral average of an observable $A$, given by
averaging the estimator $A_{\text{est}}$ over an ensemble with weight
$\rho^{(\lambda)}(\mathbf{r})$. The so-called thermodynamic
estimator $\left[  dF(\lambda)/d\lambda\right]  _{\text{th}}$ for
$dF_{P}(\lambda)/d\lambda$ is derived simply by differentiating
Eq.~(\ref{eq:Qr_PI}),
\begin{equation}
\left[  dF(\lambda)/d\lambda\right]  _{\text{th}}=-\frac{1}{\beta}\sum
_{i=1}^{N}\frac{dm_{i}}{d\lambda}\left[  \frac{DP}{2m_{i}}-\frac{P}{2\beta\hbar^{2}}
\sum_{s=1}^{P}|\mathrm{r}_{i}^{(s)}-\mathrm{r}_{i}^{(s-1)}%
|^{2}\right]  . \label{eq:F_th}%
\end{equation}
However, since it is a difference of two terms proportional to
$P$, this estimator has a statistical error that grows with the Trotter number
$P$, further increasing the computational cost. This drawback motivated the
introduction\cite{Vanicek_Miller:2007} of the centroid virial estimator
$\left[  dF(\lambda)/d\lambda\right]_{\text{cv}}$ whose
statistical error is independent of $P$, a property mirroring the property of
an analogous centroid virial estimator for kinetic
energy.\cite{Predescu_Doll:2002,Yamamoto:2005} The centroid virial
estimator, derived in Appendix~\ref{app:NMPIMC}, is given by
\begin{equation}
\left[  dF(\lambda)/d\lambda\right]  _{\text{cv}}=-\sum_{i=1}^{N}\frac
{1}{2m_{i}}\frac{dm_{i}}{d\lambda}\left\{  \frac{D}{\beta}+\frac{1}{P}%
\sum_{s=1}^{P}\left[  (\mathrm{r}_{i}^{(s)}-\mathrm{r}_{i}^{(C)})\cdot
\nabla_{i}V(\mathbf{r}^{(s)})\right]  \right\}  , \label{eq:F_cv}%
\end{equation}
where
\begin{equation}
\mathbf{r}^{(C)}:=\frac{1}{P}\sum_{s=1}^{P}\mathbf{r}^{(s)}%
\end{equation}
is the centroid coordinate of the polymer ring. All our numerical
examples use the centroid virial estimators, unless explicitly mentioned otherwise.

To summarize, using thermodynamic integration, the isotope effect
(\ref{eq:IE_definition}) is evaluated as
\begin{equation}
\frac{Q_{P}^{(B)}}{Q_{P}^{(A)}}=\exp\left\{  -\beta\int_{0}^{1}\left\langle
\left[  dF(\lambda)/d\lambda\right]  _{\text{cv}}\right\rangle ^{(\lambda
)}d\lambda\right\}  . \label{eq:thermodynamic_integration_final_formula}%
\end{equation}
The calculation of the isotope effect requires running simulations at
different values of $\lambda$ and then numerically evaluating the integral in
Eq.~(\ref{eq:thermodynamic_integration_final_formula}) using, for example, the
trapezoidal, midpoint, or Simpson rule.

%\subsection{Thermodynamic integration in extended configuration space}

\subsection{Stochastic thermodynamic integration with respect to mass}

\label{subsec:TI_in_extended_space}

It is evident that the method of thermodynamic integration introduces an
integration error, and therefore several approaches have been proposed to
decrease it: While Ceriotti and Markland\cite{Ceriotti_Markland:2013}
optimized the interpolation functions $m_{i}(\lambda)$ in order
to make $dF_{P}(\lambda)/d\lambda$ as flat as possible over the
integration interval, and thus obtained Eq.~(\ref{eq:m_root_interpol}),
Mar\v{s}\'{a}lek and Tuckerman\cite{Marsalek_Tuckerman:2014} introduced
higher-order derivatives of $Q(\lambda)$ with respect to
$\lambda$. Both modifications decrease the integration error, but do not
eliminate it completely. In this subsection we show that including
the $\lambda$ variable as an additional dimension in the Monte Carlo
simulation allows to make the integration error exactly zero if an appropriate
sampling procedure is used.

To illustrate why it makes sense to evaluate the $\lambda$ integral
stochastically, let us consider a \textquotedblleft standard\textquotedblright%
\ thermodynamic integration protocol from the previous subsection, where the
integral in Eq.~(\ref{eq:introducing_thermodynamic_integration}) is evaluated
deterministically by discretizing the $\lambda$ interval $\left[  0,1\right]
$ into $J$ subintervals of the form $I_{j}=[\lambda_{j-1},\lambda_{j}%
]$, typically with $\lambda_{j}=j/J$ ($j=0,\dots,J$). For
example, employing the midpoint rule for the integral, one would
run a separate Monte Carlo simulation for each $\bar{\lambda}_{j}:=\left(
\lambda_{j-1}+\lambda_{j}\right)  /2=\left(  j-1/2\right)  /J\in I_{j}$
($j=1,\dots,J$) in order to calculate $dF(\lambda)/d\lambda
|_{\lambda=\bar{\lambda}_{j}}$. Suppose we increase $J$ while keeping the
length of each simulation inversely proportional to $J$. Then the total number
of Monte Carlo steps used will remain constant, the integration error will
decrease, and the statistical error of the evaluated isotope effect will be
close to a limiting value as long as each individual simulation is
statistically converged. Unfortunately, for a fixed overall cost one cannot
use arbitrarily large values of $J$, since that would render the individual
simulations so short that their ergodicity would no longer be guaranteed. If
ergodicity of an individual simulation requires at least $M_{\text{erg}}$
Monte Carlo steps, the cost of the calculation will grow as $\mathcal{O}%
(J\,M_{\text{erg}})$, making the limit $J\rightarrow\infty$ unattainable in practice.

If, instead of $J$ separate simulations for each $\bar{\lambda}_{j}$, one
performs a single Monte Carlo simulation in a configuration space with an
extra dimension corresponding to $\lambda$, the average of estimator
$\left[  dF(\lambda)/d\lambda\right]  _{\text{cv}}$ over each
subinterval $I_{j}$ will give an estimate for $dF(\lambda
)/d\lambda|_{\lambda=\bar{\lambda}_{j}}$, and one can use much higher values
of $J$ (and therefore obtain smaller integration errors) without sacrificing
ergodicity of the simulation. This trick bears some resemblance to umbrella
integration\cite{Kastner_Thiel:2005,Kastner:2009,Kastner:2012}
and adaptive biasing force\cite{Darve_Pohorille:2001,Darve_Pohorille:2008,Comer_Chipot:2015} approaches
used to find the dependence of free energy on a reaction coordinate, but here
the role of reaction coordinate\ is taken by isotope masses. As in umbrella
integration, decreasing the widths of the $\lambda$ intervals $I_{j}$
decreases the integration error without affecting the statistical error of the
computed isotope effect.

Running a Monte Carlo simulation in a configuration space augmented by
$\lambda$ requires, first of all, a correct sampling weight, $\rho
^{(\lambda)}(\mathbf{r})$, which is nothing but $\rho
(\mathbf{r})$ with masses $m_{i}(\lambda)$ evaluated at a given
value $\lambda$. The second most important thing is a corresponding Monte
Carlo trial move together with an acceptance rule. The simplest
possible trial move with respect to $\lambda$
changes the initial $\lambda^{\prime}$ to any other $\lambda^{\prime\prime
}\in\lbrack0,1]$ with equal probability, and keeps the Cartesian
coordinates $\mathbf{r}$ of the ring polymer fixed.
The resulting ratio of probability densities corresponding to
$\lambda^{\prime\prime}$ and $\lambda^{\prime}$ is
\begin{equation}
\frac{\rho^{(\lambda^{\prime\prime})}(\mathbf{r})}{\rho^{(\lambda^{\prime}%
)}(\mathbf{r})}=\left[  \prod_{i=1}^{N}\frac{m_{i}(\lambda^{\prime\prime}%
)}{m_{i}(\lambda^{\prime})}\right]  ^{PD/2}\exp\left\{  \frac{P}{2\beta
\hbar^{2}}\sum_{i=1}^{N}[m_{i}(\lambda^{\prime}) -m_{i}(\lambda^{\prime\prime
})]\sum_{s=1}^{P}|\mathrm{r}_{i}^{(s)}-\mathrm{r}_{i}^{(s-1)}|^{2}\right\}  ,
\label{eq:simple_acceptance_probability}%
\end{equation}
which, as a function of $\lambda^{\prime\prime}$,
has a maximum that unfortunately becomes sharper with larger $P$.
A simple way to keep acceptance probability high even for large
values of $P $ is to generate trial $\lambda^{\prime\prime}$ such that
$\left\vert \lambda^{\prime\prime}-\lambda^{\prime}\right\vert \leq
\Delta\lambda_{{\mathrm{max}}}$. The following Monte Carlo procedure satisfies
this condition and also preserves the acceptance ratio given by Eq.
(\ref{eq:simple_acceptance_probability}):

\textbf{Simple $\lambda$-move:}

\begin{enumerate}
\setlength{\itemindent}{0.05\textwidth}

\item Trial move:
\begin{align}
\lambda^{\prime}  &  \mapsto\lambda^{\prime\prime}=\lambda^{\prime}%
+\Delta\lambda\text{, where}\label{eq:straightforward_lambda_trial_move}\\
&  \Delta\lambda\in[-\Delta\lambda_{\mathrm{max}},\Delta\lambda_{\mathrm{max}%
}] \text{ and distributed uniformly.}
\label{eq:straightforward_lambda_trial_restriction}%
\end{align}

\item Readjust the trial move to satisfy $\lambda^{\prime\prime}\in[0,1]$:
\begin{align}
\text{if }(\lambda^{\prime\prime}<0)\text{ then }  &  \lambda^{\prime\prime
}\mapsto-\lambda^{\prime\prime},\\
\text{if }(\lambda^{\prime\prime}>1)\text{ then }  &  \lambda^{\prime\prime
}\mapsto2-\lambda^{\prime\prime}. \label{eq:trial_lambda_move_readjustment}%
\end{align}

\item Accept the final trial move with a probability%
\begin{equation}
\mathrm{min}\left(  1,\left[  \prod_{i=1}^{N} \frac{m_{i}(\lambda
^{\prime\prime})}{m_{i}(\lambda^{\prime})}\right]  ^{PD/2}\exp\left\{
\frac{P}{2\beta\hbar^{2}}\sum_{i=1}^{N}[m_{i}(\lambda^{\prime})-m_{i}%
(\lambda^{\prime\prime})]\sum_{s=1}^{P}|\mathrm{r}_{i}^{(s)}-\mathrm{r}%
_{i}^{(s-1)}|^{2}\right\}  \right)  .
\label{eq:straightforward_lambda_MC_procedure}%
\end{equation}

\end{enumerate}

The procedure defined by Eqs.~(\ref{eq:straightforward_lambda_trial_move}%
)-(\ref{eq:straightforward_lambda_MC_procedure}) is almost free in terms of
computational time, but at very large values of $P $, even with the
restriction~(\ref{eq:straightforward_lambda_trial_restriction}), it becomes
ineffective at sampling $\lambda$ values far from the maximum of the
probability ratio~(\ref{eq:simple_acceptance_probability}). This problem can
be bypassed if the trial move with respect to $\lambda$ preserves the
mass-scaled normal modes of the ring polymer instead of the Cartesian
coordinates, resulting in the following Monte Carlo procedure derived in
Appendix~\ref{app:NMPIMC}:

\textbf{Mass-scaled $\lambda$-move:}

\begin{enumerate}
\setlength{\itemindent}{0.05\textwidth}

\item Trial move:%
\begin{align}
\lambda^{\prime}  &  \mapsto\lambda^{\prime\prime}\in\left[  0,1\right]
\text{ and distributed uniformly},\label{eq:lambda_trial_move}\\
\mathbf{r}  &  \mapsto\mathbf{r}_{\lambda^{\prime},\lambda^{\prime\prime}},
\end{align}
where%
\begin{equation}
\mathrm{r}_{\lambda^{\prime},\lambda^{\prime\prime},i}^{(s)}:=\mathrm{r}%
_{i}^{(C)}+\sqrt{\frac{m_{i}(\lambda^{\prime})}{m_{i}(\lambda^{\prime\prime}%
)}}(\mathrm{r}_{i}^{(s)}-\mathrm{r}_{i}^{(C)})\text{.}
\label{eq:rescaled_r_defined}%
\end{equation}

\item Accept the trial move with a probability%
\begin{equation}
\mathrm{min}\left(  1,\left[  \prod_{i=1}^{N}\frac{m_{i}(\lambda^{\prime
\prime})}{m_{i}(\lambda^{\prime})}\right]  ^{D/2}\exp\left\{  \frac{\beta}%
{P}\sum_{s=1}^{P}[V(\mathbf{r}^{(s)})-V(\mathbf{r}_{\lambda^{\prime}%
,\lambda^{\prime\prime}}^{(s)})]\right\}  \right)  .
\label{eq:lambda_MC_procedure}%
\end{equation}

\end{enumerate}

When discussing Monte Carlo moves with respect to $\lambda$, we shall refer
the procedure defined by Eqs.~(\ref{eq:straightforward_lambda_trial_move}%
)-(\ref{eq:straightforward_lambda_MC_procedure}) as the ``simple $\lambda
$-move'', and to that of Eqs.~(\ref{eq:lambda_trial_move}%
)-(\ref{eq:lambda_MC_procedure}) as the ``mass-scaled $\lambda$-move''. If the
centroid probability distribution starts to vary too much over $\lambda
\in[0,1]$, the acceptance probability for the mass-scaled $\lambda$-move can
become too low; this is solved easily by restricting the trial $\lambda
^{\prime\prime}$ value to a smaller interval $[\lambda^{\prime}-\Delta
\lambda_{\mathrm{max}},\lambda^{\prime}+\Delta\lambda_{\max}]$ using the
procedure of Eqs.~(\ref{eq:straightforward_lambda_trial_move}%
)-(\ref{eq:trial_lambda_move_readjustment}). [Yet, for all systems considered
in this work, Eqs.~(\ref{eq:lambda_trial_move})-(\ref{eq:lambda_MC_procedure})
led to sufficiently high acceptance probability without this modification.]
The main advantage of the mass-scaled $\lambda$-move is that its acceptance
probability does not depend on $P$. Its disadvantage is its requirement of $P$
evaluations of $V$, which makes it much more expensive than the simple
$\lambda$-move. Nonetheless, as will be demonstrated in
Sec.~\ref{seq:Applications}, an occasional use of mass-scaled $\lambda$-moves
can, in fact, accelerate convergence with respect to $\lambda$.

The Monte Carlo procedure has one last shortcoming: Since the
probability of finding the system with $\lambda=\lambda^{\prime}$ is
proportional to $Q(\lambda^{\prime})$, for very large isotope
effects (the largest isotope effect computed in this work was $\sim10^{8}$)
most of the samples would be taken in the region close to $\lambda=0$, which
would introduce a huge statistical error. This problem can be solved by adding
a biasing umbrella potential $U_{b}(\lambda)$, resulting in a
biased probability density
\begin{equation}
\rho_{b}^{(\lambda)}(\mathbf{r})=\rho^{(\lambda)}(\mathbf{r})\exp[-\beta
U_{b}(\lambda)]. \label{eq:biased_extended_probability_density}%
\end{equation}
In the case of a free particle, all trial moves defined by
Eqs.~(\ref{eq:lambda_trial_move})-(\ref{eq:rescaled_r_defined})
will be accepted provided that the optimal biasing potential
\begin{equation}
U_{b,\mathrm{free}}(\lambda)=\frac{D}{2\beta}\sum_{i=1}^{N}\ln[m_{i}%
(\lambda)/m_{i}(0)]
\end{equation}
is chosen; in other words, if $V\equiv0$, then including
$U_{b,\mathrm{free}}(\lambda)$ in the acceptance probability
(\ref{eq:lambda_MC_procedure}) will make it unity.

With this final modification in place, the proposed method can be summarized
as running a Monte Carlo simulation in the augmented configuration space and
then evaluating the isotope effect with the formula
\begin{equation}
\frac{Q_{P}^{(B)}}{Q_{P}^{(A)}}=\lim_{J\rightarrow\infty}\exp\left[
-\frac{\beta}{J}\sum_{j=1}^{J}\left\langle \left[  dF(\lambda)/d\lambda
\right]  _{\text{cv}}\right\rangle ^{I_{j}}\right]  ,
\label{eq:lambda_TI_final_formula}%
\end{equation}
where $\langle\cdots\rangle^{I_{j}}$ is an average over all $\lambda\in I_{j}%
$. The integration error associated with having a finite number
$J$ of $\lambda$ intervals depends strongly on the choice of the umbrella
potential $U_{b}(\lambda)$. As we prove in
Appendix~\ref{app:STI_error_control}, this error is exactly zero for a
piecewise linear umbrella potential satisfying
\begin{equation}
\frac{dU_{b}(\lambda)}{d\lambda}=-\langle\lbrack dF(\lambda)/d\lambda
]_{\mathrm{cv}}\rangle^{I_{j}}\text{ for all $\lambda\in I_{j}$}.
\label{eq:U_b_choice}%
\end{equation}
It is also clear that the resulting $U_{b}(\lambda)$ will follow fairly
closely the ideal biasing potential $\beta^{-1}\ln Q(\lambda)$, and therefore
the estimator samples will be distributed more or less equally among different
intervals $I_{j}$, which, in turn, will minimize the statistical error of
Eq.~(\ref{eq:lambda_TI_final_formula}).

It is obvious that in general systems, $U_{b}(\lambda)$ from
Eq.~(\ref{eq:U_b_choice}) cannot be known \textit{a priori}. As this is
typical for biased simulations, numerous methods, including adaptive umbrella
sampling,\cite{Mezei:1987, Hooft_Kroon:1992,Bartels_Karplus:1997}
metadynamics,\cite{Micheletti_Parrinello:2004,Laio_Parrinello:2005} and
adaptive biasing force method,\cite{Darve_Pohorille:2001,Darve_Pohorille:2008}
have been introduced to solve this problem. In our calculations, the biasing
potential $U_{b}(\lambda)$ was obtained from a short simulation employing the
adaptive biasing force method. The resulting $U_{b}(\lambda)$ was then used in
a longer simulation in which the isotope effect itself was evaluated.

\section{Numerical examples}

\label{seq:Applications}

In this section the proposed stochastic procedure for evaluating
isotope effects is tested on a model harmonic system and
on deuteration of methane. The results of the new approach
are compared with results of the usual thermodynamic integration
and with the analytical result for the harmonic system. From now
on, for brevity we will refer to the traditional thermodynamic
integration with respect to mass (Subsec.~\ref{subsec:TI}) simply as
\textquotedblleft thermodynamic integration\textquotedblright\ (TI), and to
the thermodynamic integration with stochastic change of mass
(Subsec.~\ref{subsec:TI_in_extended_space}) as \textquotedblleft stochastic
thermodynamic integration\textquotedblright\ (STI). In all cases,
we compare STI with TI both for the linear
[Eq.~(\ref{eq:m_linear_interpol})] and the more efficient
[Eq.~(\ref{eq:m_root_interpol})] interpolation of mass.

\subsection{Computational details}

\label{subsec:global_comp_details}

As mentioned in Sec.~\ref{sec:Theory}, the $\lambda$ interval $\left[
0,1\right]  $ was divided into $J$ subintervals $I_{j}=[\lambda_{j-1}%
,\lambda_{j}]$ $\left(  j=1,\ldots,J\right)  $ with $\lambda_{j}=j/J$
($j=0,\dots,J$). The TI used, in addition, a reference point $\bar{\lambda
}_{j}$ from each interval, which was always taken to be the midpoint
$\overline{\lambda}_{j}=(j-1/2)/J\in I_{j}$. This midpoint was used for
evaluating the thermodynamic integral with the midpoint rule as%
\begin{equation}
\int_{0}^{1}\frac{d\ln Q(\lambda)}{d\lambda}d\lambda=\frac{1}{J}\sum_{j=1}%
^{J}\left.  \frac{d\ln Q(\lambda)}{d\lambda}\right\vert _{\lambda=\bar
{\lambda}_{j}}+\mathcal{O}\left(  J^{-2}\right)  . \label{eq:TI_int_error}%
\end{equation}
(Assuming that each logarithmic derivative is obtained with the same
statistical error, this choice of $\bar{\lambda}_{j}$'s and integration scheme
minimizes the statistical error of the logarithm of the
calculated isotope effect.) To estimate the integration error of
TI and to verify that the integration error of STI is zero, we compared the
calculated isotope effects with the exact
analytical\cite{Schweizer_Wolynes:1981} values for the harmonic system with a
finite Trotter number $P$ and with the result of STI using a high value of
$J=8192$ for the deuteration of methane.

The second type of error is the statistical error inherent to all Monte Carlo
methods; this error was evaluated with the \textquotedblleft
block-averaging\textquotedblright\ method\ \cite{Flyvbjerg_Petersen:1989} for
correlated samples, which was applied directly to the computed isotope effects
instead of, e.g., the free energy derivatives, thus avoiding the tedious error
propagation. Since the average isotope effect depends on the block size, one
has to make sure not only that the statistical error reaches a plateau, but
also that the average reaches an asymptotic value as a function of the block size.

The third type of error is the Boltzmann operator discretization error due to
a finite value of $P$; for harmonic systems it is available
analytically,\cite{Schweizer_Wolynes:1981} while for the $\mathrm{CD_{4}%
/CH_{4}}$ isotope effect we made sure that it was below $1\%$ by repeating the
calculations for the lowest and highest temperatures with twice larger $P$.

\subsection{\label{subsec:harmonic}Isotope effects in a harmonic model}

A harmonic system was used as the first, benchmark test of the different
approaches to compute the isotope effects, since most properties
of a harmonic system can be computed exactly analytically. To simulate a
realistic system with a range of vibrational frequencies, we used an
eight-dimensional harmonic system with frequencies
\begin{equation}
\omega_{q}=\omega_{0}\times2^{-q/2}\text{ \ \ }(q=0,\dots,7).
\label{eq:omega_q_choice}%
\end{equation}
The computed isotope effect corresponded to doubling masses of all normal
modes, and therefore to reducing each $\omega_{q}$ by a factor of $\sqrt{2}$.

\subsubsection{Computational details}

To analyze the dependence of the computed isotope effect on the number $J$ of
$\lambda$ intervals used in different methods, we first ran several
calculations with $\beta\hbar\omega_{0}=8$. Then we investigated
the behavior of the different methods at several temperatures and
hence for dramatically different isotope effects, by taking $\beta\hbar
\omega_{0}\in\{1,2,4,8,16,32\}$ (here we used $J=8$ for TI and $J=4096$ for
STI). For each $\omega_{0}$ the Trotter number $P$ was chosen so that the
discretization error of the isotope effect (i.e., not of its logarithm) was
below $1\%$.

To explore the ring polymer coordinates $\mathbf{r}$, we used the
normal mode path integral Monte Carlo
method,\cite{Herman_Berne:1982,Cao_Berne:1993} which in a harmonic model
allows to generate uncorrelated samples with no rejected Monte Carlo steps.
This method involves rewriting $\Phi(\mathbf{r})$ in terms of
normal modes of the ring polymer (see Appendix~\ref{app:NMPIMC}),
thus transforming $\rho(\mathbf{r})$ into a product of Gaussians
that can be sampled exactly. In all TI calculations
the total number of Monte Carlo steps was $2^{25}\approx3.4\times10^{7}$. In
all STI calculations we used a mixture of $9\times2^{22}\approx3.7\times
10^{7}$ Monte Carlo moves with respect to $\mathbf{r}$, $2^{22}\approx
4.2\times10^{6}$ mass-scaled $\lambda$-moves, and $5\times2^{23}%
\approx4.2\times10^{7}$ simple $\lambda$-moves with $\Delta\lambda
_{\mathrm{max}}=0.1$; first $20\%$ of a STI calculation were used only to
obtain the biasing potential $U_{b}(\lambda)$, but not for evaluating the IE.
Note that the unequal numbers of Monte Carlo steps used in TI and STI result
in a fair comparison of the two methods; the simple $\lambda$-moves are almost
free in terms of computational effort, and, due to warmup, the total number of
the other Monte Carlo moves for STI is $20\%$ larger than for TI, which is not
an issue, since generally (i.e., in anharmonic systems in which the sampling
procedure would generate correlated samples) one would need to discard a
certain warmup period also in TI calculations.

\subsubsection{Results and discussion}

\label{subsubsec:ho_num_res}

\begin{figure}
[ptbh]\centering\includegraphics[width=\textwidth]{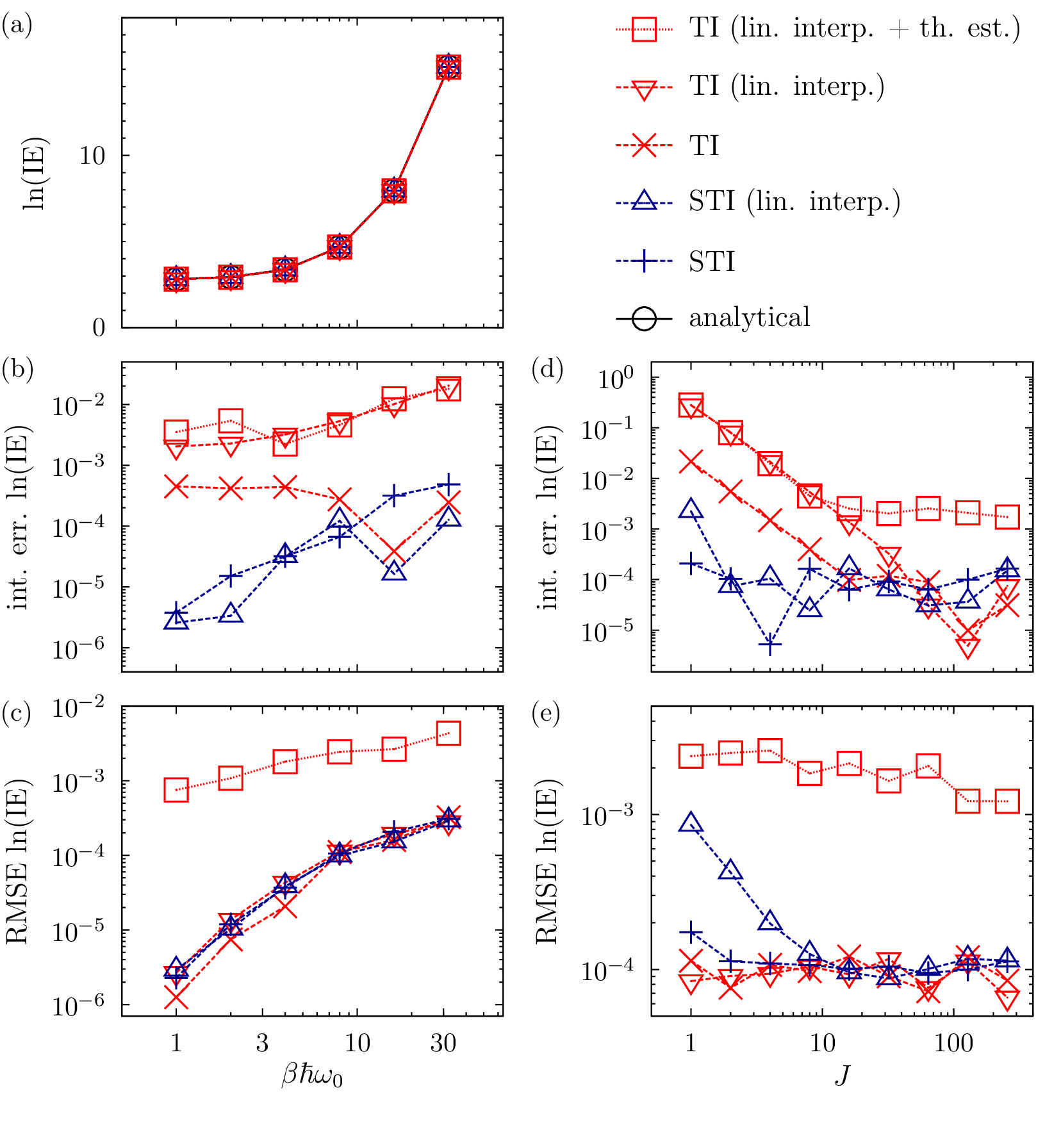}
\caption{\label{fig:ho_num_performance}Isotope effect (IE) calculations in an eight-dimensional harmonic model
from Subsec.~\ref{subsec:harmonic}. Unless explicitly stated in the label, all results use the centroid
virial estimator~(\ref{eq:F_cv}) and improved mass interpolation~(\ref{eq:m_root_interpol}). Results labeled
``lin.~interp.'' use linear interpolation~(\ref{eq:m_linear_interpol}) and those labeled ``th.~est.'' the
thermodynamic estimator~(\ref{eq:F_th}). Several versions of thermodynamic integration (TI) are compared
with exact analytical values (for the same finite Trotter number $P$). The proposed method is
``stochastic thermodynamics integration'' (STI). Panels (a)-(c) show the temperature dependence of
(a) the isotope effect,
(b) its integration errors, and
(c) its statistical root mean square errors (RMSEs).
Panels (d)-(e) display the dependence of integration errors and RMSEs on the number $J$ of
integration subintervals at a temperature given by $ \beta\hbar\omega_{0}=8$.
}
\end{figure}

The numerical results are presented in Fig.~\ref{fig:ho_num_performance}.
Panel~(a) of the figure shows that analytical values of the
isotope effect (at a finite value of $P$) are reproduced accurately by STI for
several values of $\beta\hbar\omega_{0}$, confirming that the proposed Monte
Carlo procedure, which changes stochastically not only coordinates but also
masses of the atoms, is correct.

Panel~(b) displays the integration error dependence on
temperature, and confirms that this error is decreased both by linearly
interpolating the inverse square roots of the masses instead of the masses
themselves, and by performing the thermodynamic integration stochastically.
The fact that the stochastic change of mass can eliminate the thermodynamic
integration error is the main result of this paper. As the figure shows, this
happens regardless of the type of interpolation used. Note that at
high temperatures, the improved interpolation does not prevent TI from
exhibiting a certain integration error, an issue that does not occur for STI.
The statistical error dependence on temperature, depicted in panel (c),
is a reminder of the well-known importance of using
the centroid virial instead of the thermodynamic
estimator in efficient calculations. In the harmonic system,
which can be sampled exactly, the statistical errors of STI and TI are comparable.

Panels (d) and (e) of Fig.~\ref{fig:ho_num_performance} display
the dependence of integration and statistical errors of different methods on
the number $J$ of integration subintervals for $\beta\hbar\omega_{0}=8$. For
TI one can clearly see the $J\rightarrow\infty$ limit where integration error
becomes zero and statistical error approaches a plateau. Note that the
integration error [panels (b) and (d)] does not depend on the estimator, which
provides an additional check of the implementation. The centroid virial
estimator significantly lowers the statistical error and using the square root
of mass interpolation given by Eq.~(\ref{eq:m_root_interpol}) instead of
linear interpolation [Eq.~(\ref{eq:m_linear_interpol})] significantly
decreases the integration error. As expected, STI exhibits an error which is
only due to statistical factors. Here the TI and STI exhibit similar behavior
in the $J\rightarrow\infty$ limit, namely the integration error is zero and
the statistical error approaches a limit which is comparable for both methods.
However, in this system the limit $J\rightarrow\infty$ was achievable for TI
because the normal mode path integral Monte Carlo procedure used for exploring
$\mathbf{r}$ generated uncorrelated samples; reaching $J\rightarrow\infty$
would be more difficult in more realistic, anharmonic systems, where even the
TI procedure requires correlated sampling. Yet, as will be shown below on
methane, large values of $J$ can be used easily in STI calculations. Also note
that the statistical error of STI decreases with $J$ and approaches its limit
faster when the square root of mass interpolation
[Eq.~(\ref{eq:m_root_interpol})] is used. This behavior is expected as the
statistical error of $\langle\lbrack dF(\lambda)/d\lambda]_{\mathrm{cv}%
}\rangle^{I_{j}}$ from Eq.~(\ref{eq:lambda_TI_final_formula}) is partly due to
a variation of the average $\langle\lbrack dF(\lambda)/d\lambda]_{\mathrm{cv}%
}\rangle^{(\lambda)}$ over $\lambda\in I_{j}$; the resulting contribution to
the statistical error of IE is reduced by increasing $J$ or using an improved
mass interpolation function that makes $\langle\lbrack dF(\lambda
)/d\lambda]_{\mathrm{cv}}\rangle^{(\lambda)}$ flatter over each $I_{j}$.

\subsection{Deuteration of methane}

\label{subsec:methane_deuterization}

\subsubsection{Computational details}

The methane calculations used the potential energy surface from
Ref.~\onlinecite{Schwenke_Partridge:2001} and available in the POTLIB
library.\cite{POTLIB} The number of $\lambda$ integration
intervals was $J=4$ for TI and $J=4096$ for STI.

TI calculations used a total of $2\times10^{8}$ Monte Carlo steps
which sampled $\mathbf{r}$, for STI the number of $\mathbf{r}$ Monte Carlo
steps was $1.8\times10^{8}$; in both cases {$14\%$} were whole-chain moves and
$86\%$ were multi-slice moves performed on one sixth of the chain with the
staging algorithm\cite{Sprik_Chandler:1985,Sprik_Chandler:1985_1} (this
guaranteed that approximately the same computer time was spent on either of
the two types of moves). For STI we additionally used $0.2\times10^{8}$
mass-scaled $\lambda$-moves and $2\times10^{8}$ simple $\lambda$-moves with
$\Delta\lambda_{\mathrm{max}}=0.1$. As the simple $\lambda$-moves are almost
free in terms of computational time, the cost of both calculations was still
roughly the same. To avoid the unnecessary cost of evaluating correlated
samples, all virial estimators were evaluated only after every ten Monte Carlo
steps for TI and after every twenty Monte Carlo steps for STI (since STI
calculations had twice as many Monte Carlo steps the number of virial
estimator samples was still the same), while thermodynamic estimators were
evaluated after each step since the CPU time required for their calculation is
negligible. The first $20\%$ Monte Carlo steps of each calculation were
discarded as \textquotedblleft warmup\textquotedblright; as discussed in
Subsec.~\ref{subsec:global_comp_details}, in the simulations employing the
stochastic change of mass, the same warmup\ period was also used to generate
the biasing potential $U_{b}(\lambda)$ needed for the rest of the calculation.
The path integral discretization error, estimated by running simulations with
a twice larger $P$ at the highest and lowest temperatures ($T=1000\,$K and
$T=200\,$K), was below $1\%$; for other temperatures $P$ was obtained by
linear interpolation with respect to $1/T$.

Of course, in practice much shorter simulations would be sufficient, but we
used overconverged calculations in order to analyze the behavior of different
types of errors in detail.

\subsubsection{Results and discussion}

\begin{figure}
[tbp]\centering\includegraphics[width=\textwidth]{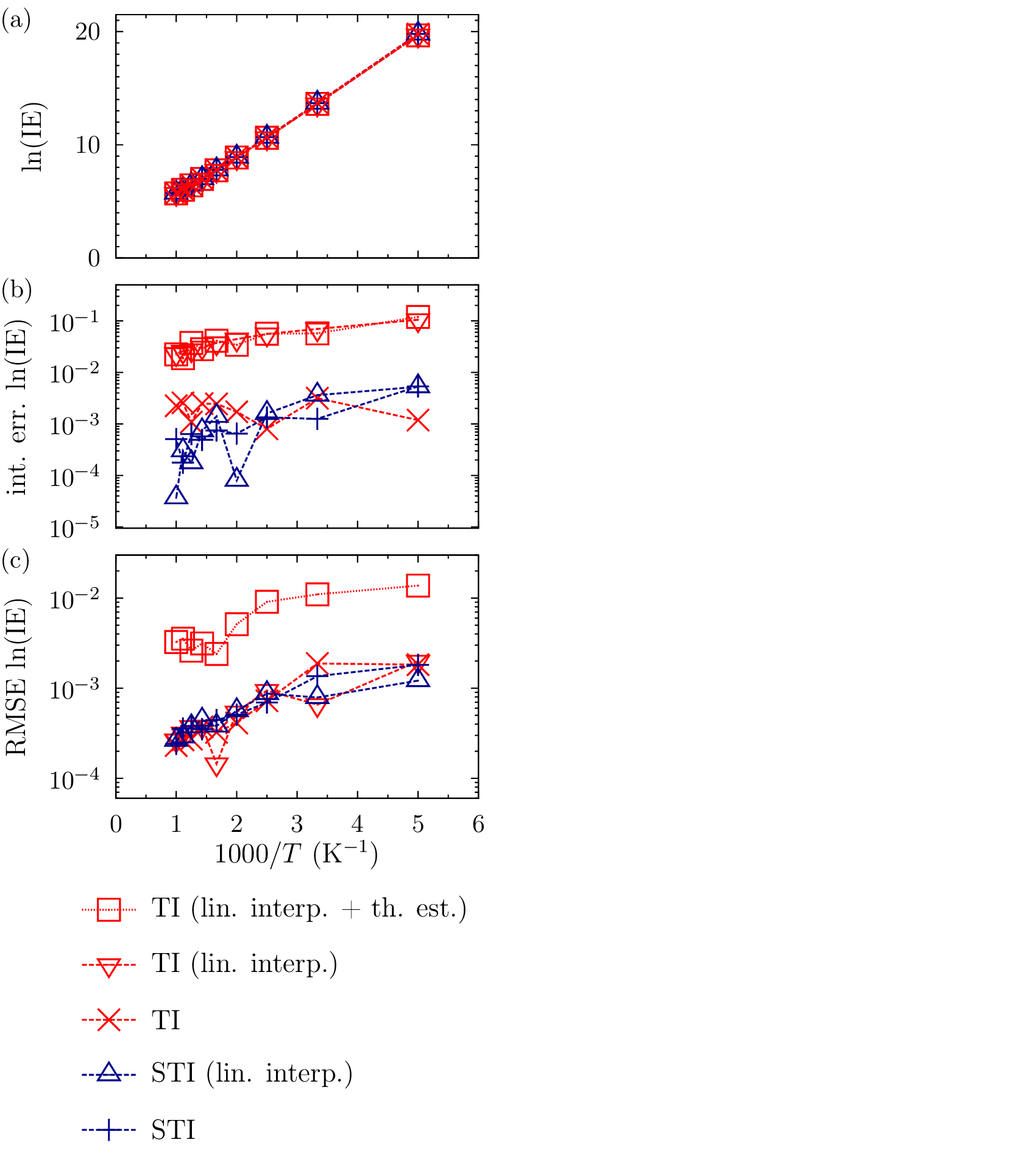}
\caption{\label{fig:ch4_num_performance} Calculations of the $\mathrm{CD_{4}/CH_{4}}$ isotope effect (IE) from
Subsec.~\ref{subsec:methane_deuterization}.
Labels are explained in the caption of Fig.~\ref{fig:ho_num_performance}. The three panels show
the temperature dependence of (a) the  isotope effect, (b) its integration errors, and (c) its statistical root
mean square errors (RMSEs).}
\end{figure}

The results of the calculations of the $\mathrm{CD}_{4}/\mathrm{CH}_{4}$
isotope effect are presented in Fig.~\ref{fig:ch4_num_performance}.
Panel (a) shows that the isotope effects calculated with the
different methods agree. Yet, a more detailed inspection reveals the
improvement provided by the STI compared with the TI. This is done in
panel~(b), showing the integration errors of the different
methods; the STI result with a twice larger value of $J$ (i.e.,
$J=8192$) is considered as an exact benchmark. In the case of TI
the integration error depends strongly on the type of mass interpolation: If
the linear interpolation is used, the integration error is even much larger
than the statistical error [see panel (c)], while, for this particular system,
the improved interpolation (\ref{eq:m_root_interpol}) of the inverse square
root of mass allows to obtain quite accurate results, even though a small
integration error remains visible above the statistical noise at higher
temperatures. In the case of the STI, on the other hand, no
integration error is observed, which was one of the main goals of this work.
Finally, panel~(c) shows that if the same estimator
is used the STI exhibits comparable statistical errors to those of TI, which
confirms that employing the STI can easily lower the integration errors
without increasing the computational cost.

For reference, the plotted values together with their statistical
errors are listed in Table~\ref{tab:lnIE_values_and_disc_error}. From this
table it is clear that STI calculations\ with both types of mass interpolation
agree within their statistical errors, while TI, particularly with linear
interpolation, retains a significant integration error.

\begin{table}
\caption{Values of the $ \mathrm{CD}_{4}/\mathrm{CH}_{4} $ isotope effect (IE) obtained with several versions of
thermodynamic integration (TI). Corresponding statistical errors are shown as well.
Unless explicitly stated in the label, all results use the centroid virial estimator (\ref{eq:F_cv}) and improved mass
interpolation (\ref{eq:m_root_interpol}). The proposed methodology is stochastic thermodynamic integration (STI).
\hfill}\label{tab:lnIE_values_and_disc_error} \begin{ruledtabular}
\begin{tabular}{ccccccc}
\multirow{2}{*}{$ T $} & \multirow{2}{*}{$ P $} & \multicolumn{5}{c}{ln(IE) ($\mathrm{CD_{4}/CH_{4}}$) with statistical error}\tabularnewline
\cline{3-7}
&  & TI (lin. interp.   & \multirow{2}{*}{TI (lin. interp.)}& \multirow{2}{*}{TI} & \multirow{2}{*}{STI (lin. interp.)} &
\multirow{2}{*}{STI} \tabularnewline
&  & + thermod. est.)  & & & & \tabularnewline
\hline
200 & 360 & $19.67\phantom{0}\pm0.02\phantom{0}$ & $ 19.683\pm 0.002$ & $ 19.785\pm 0.002$ & $19.783\pm 0.002 $ &
$ 19.789\pm0.002$\tabularnewline
300 & 226 & $13.61\phantom{0}\pm0.01\phantom{0}$ & $ 13.602\pm 0.001$ & $ 13.671\pm 0.002$ & $ 13.676\pm 0.001$ &
$ 13.673 \pm0.002$\tabularnewline
400 & 158 & $10.61\phantom{0}\pm0.01\phantom{0}$ & $ 10.612\pm 0.001$ & $ 10.665\pm 0.001$ & $ 10.666\pm 0.001$
& $ 10.667\pm0.001 $\tabularnewline
500 & 118 & $\phantom{0}8.876\pm0.005$ & $ \phantom{0}8.866\pm 0.001$ & $ \phantom{0}8.908\pm 0.001$ & $ \phantom{0}8.910\pm 0.001$
& $ \phantom{0}8.910\pm 0.001 $\tabularnewline
600 & 90 & $\phantom{0}7.740\pm0.003$ & $ \phantom{0}7.743\pm 0.001$ & $ \phantom{0}7.778\pm 0.001$ & $ \phantom{0}7.779\pm 0.001 $
& $ \phantom{0}7.779\pm 0.001$\tabularnewline
700 & 72 & $\phantom{0}6.977\pm0.003$ & $ \phantom{0}6.974\pm 0.001$ & $ \phantom{0}7.004\pm 0.001$ & $ \phantom{0}7.006\pm 0.001 $
& $ \phantom{0}7.006\pm 0.001 $\tabularnewline
800 & 58 & $\phantom{0}6.413\pm0.003$ & $ \phantom{0}6.423 \pm 0.001$ & $ \phantom{0}6.449 \pm 0.001$ & $ \phantom{0}6.451\pm 0.001$
& $ \phantom{0}6.451\pm0.001 $\tabularnewline
900 & 46 & $\phantom{0}6.020\pm0.003$ & $ \phantom{0}6.013\pm 0.001$ & $ \phantom{0}6.037\pm 0.001$ & $ \phantom{0}6.039\pm 0.001$
& $ \phantom{0}6.039\pm0.001 $\tabularnewline
1000 & 36 & $\phantom{0}5.703\pm0.003$ & $ \phantom{0}5.702\pm 0.001$ & $ \phantom{0}5.723\pm 0.001$ & $\phantom{0}5.725\pm 0.001 $
& $ \phantom{0}5.726\pm0.001 $\tabularnewline
\end{tabular}
\end{ruledtabular}

\end{table}

To better understand the benefit of the STI, recall\textbf{ }that
stopping a Monte Carlo simulation after obtaining only a finite number of
samples introduces two types of errors. The first is the statistical error,
which has been analyzed in all calculations so far; the second type is a
systematic error, and appears if the sampling procedure yields correlated
samples and Monte Carlo trajectories are too short to guarantee ergodicity.
This systematic error has not appeared yet since all our calculations were too
well converged; however, it becomes important when computational resources are
limited, and therefore deserves additional consideration. Indeed, one of the
main motivations behind this work was the expectation that equilibrating a
single STI simulation should consume fewer computational resources than
equilibrating $J$ simulations required in a standard TI calculation. To
illustrate this point we ran several much less converged calculations of the
$\mathrm{CD}_{4}/\mathrm{CH}_{4}$ isotope effect at $T=200$ K. The number of
Monte Carlo steps used during the simulations was doubled from one calculation
to the next; for example, for TI there were $1280,2560,...,1310720$ Monte
Carlo steps partitioned in the same way as for the more converged
calculations. The only difference was that this time no part of the simulation
was discarded as warmup. Moreover, to be sure that the error observed for
smaller numbers of Monte Carlo steps is the systematic error due to
non-ergodicity and not the \textquotedblleft true\textquotedblright%
\ statistical error, the value obtained with $1280$ Monte Carlo steps was
averaged over $4096$ independent calculations, $2560$ - over $2048$
calculations, etc.; this averaging ensured that each result had roughly the
same statistical error. The STI calculations were performed with or without
the mass-scaled $\lambda$-moves and with or without the simple $\lambda$-moves
to compare the efficiency of the resulting methods.

\begin{figure}
[ptbh]%
\centering\includegraphics[width=\textwidth]{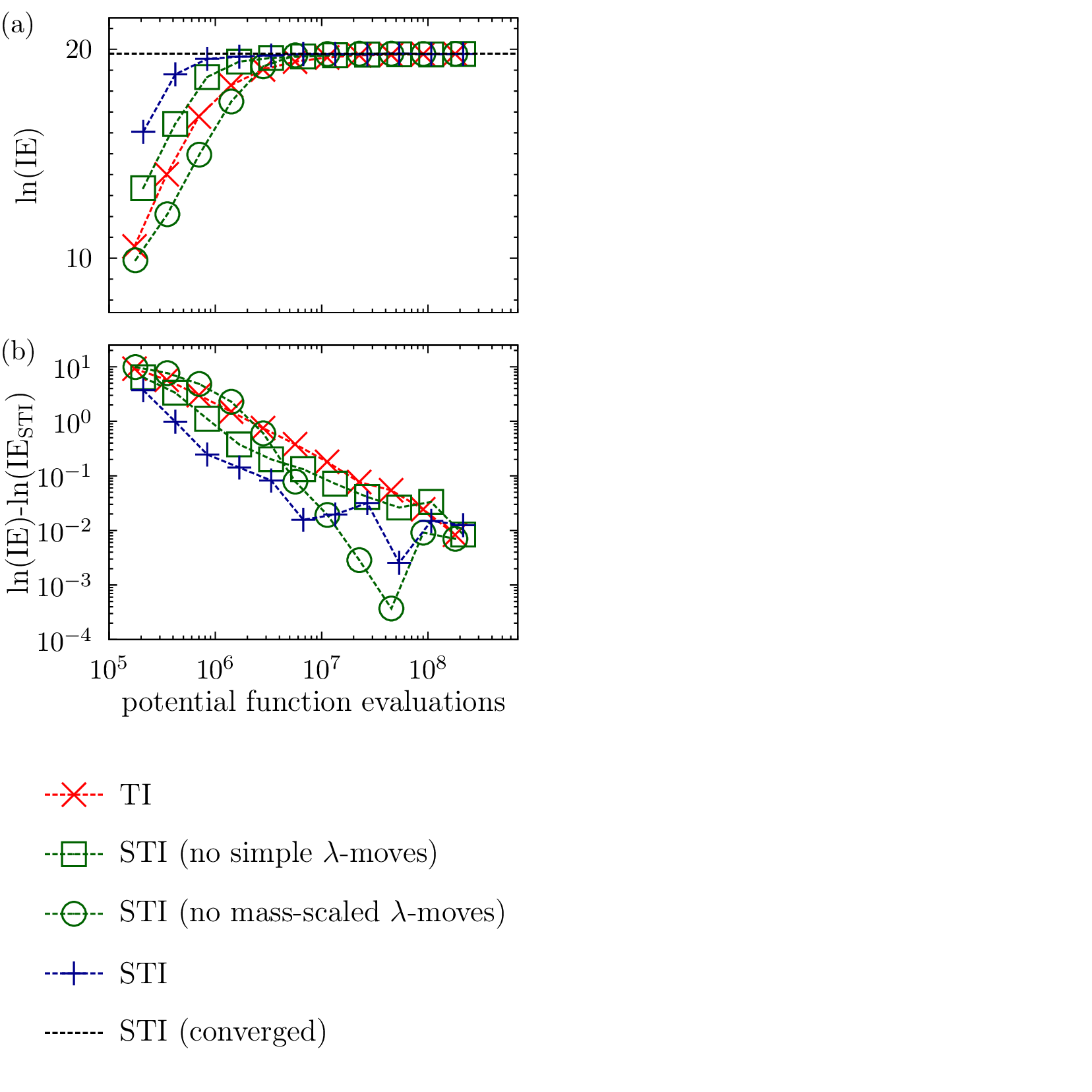}
\caption{\label{fig:ch4_nonergodicity}
The impact of nonergodicity appearing in shorter calculations of the $ \mathrm{CD}_{4}/\mathrm{CH}_{4}$ isotope effect
(IE) at $T=200$ K. Panel
(a) presents the convergence of the IE as a function of the simulation length, while panel (b) shows the corresponding error
of the IE (in logarithmic scale) relative to a converged STI result.
Labels TI and STI are as in the caption of Fig.~\ref{fig:ho_num_performance};
``STI (no simple $ \lambda$-moves)'' were obtained without the simple $ \lambda$-moves
defined by Eqs.~(\ref{eq:straightforward_lambda_trial_move})-(\ref{eq:straightforward_lambda_MC_procedure}),
while ``STI (no mass-scaled $ \lambda$-moves)'' were obtained without mass-scaled $ \lambda$-moves defined by Eqs.~(\ref{eq:lambda_trial_move})-(\ref{eq:lambda_MC_procedure}).
The horizontal line in panel (a) labeled ``STI (converged)'' is the converged STI result
$\ln(\text{IE}_{\mathrm{STI}})=19.789$ from Fig.~\ref{fig:ch4_num_performance} and Table~\ref{tab:lnIE_values_and_disc_error};
the same value was used as a reference in panel (b).}
\end{figure}

Isotope effects obtained with these much cheaper calculations are compared in
Fig.~\ref{fig:ch4_nonergodicity}, where the converged STI result $\ln
\text{IE}=19.789$ from Fig.~\ref{fig:ch4_num_performance} and
Table~\ref{tab:lnIE_values_and_disc_error} serves as the exact reference; for
a completely fair comparison the results are plotted as a function of the
number of potential energy evaluations required to obtain them. As expected,
the results of shorter simulations exhibit a significant error due to
non-ergodicity of underlying simulations, yet this nonergodicity error is
much smaller for the proposed STI than for the TI, making
the STI more practical in situations where computational
resources are limited. Even though the
mass-scaled $\lambda$-moves are quite expensive, their addition accelerates
the convergence of the integral. The much cheaper simple $\lambda$-moves
appear to also contribute to convergence, as the results
obtained without them are not as well converged as results with both types of
$\lambda$-moves.

\section{Conclusion}

We have introduced a new Monte Carlo procedure that involves
changing atomic masses stochastically during the simulation and allows to
eliminate the integration error of thermodynamic integration, thus significantly
speeding up isotope effects calculations. The proposed methodology relies on a
set of new tools: One of these tools is the introduction of mass-scaled
$\lambda$-moves that permit drastic changes of $\lambda$ in a single Monte
Carlo step; as shown in Subsec.~\ref{subsec:methane_deuterization} their
addition can significantly contribute to the convergence of the thermodynamic
integral. Another tool is the piecewise linear umbrella biasing potential
$U_{b}(\lambda)$ that guarantees a zero integration error of the thermodynamic
integral for any number $J$ of integration subintervals; this trick is general
and can be used regardless of the type of free energy change one may want to evaluate.

It is possible, as in metadynamics, to facilitate convergence with respect to
$\lambda$ by additionally biasing the simulation with a history-dependent
potential that pushes the system into less explored regions of configuration
space; this addition can become important if the change of isotope masses
$m_{i}(\lambda)$ is so drastic that one has to impose an upper bound
$\Delta\lambda_{\mathrm{max}}$ for the change of $\lambda$ in a single step
even for the mass-scaled $\lambda$-moves. However, this did not occur in
systems considered in this work, where mass-scaled $\lambda$-moves yielded
acceptance probabilities above $70 \%$ in all calculations. As a result, the
mass-scaled $\lambda$-moves allowed large changes of $\lambda$ in a single
step, leading to a fast convergence over the $\lambda$ dimension without
additional modifications of the Monte Carlo procedure.

In this work we relied heavily on the fact that $\lambda$ values can be
sampled without repercussions even if they are placed far away from the
endpoints $\lambda=0$ and $\lambda=1$ that correspond to physically meaningful
systems. This is true for IEs, as also found in
Ref.~\onlinecite{Perez_Lilienfeld:2011}, but may not be so for other
calculations of free energy differences. As a result, several variants of
$\lambda$-dynamics bias the sampling of $\lambda$ towards the endpoints, and
then calculate the free energy difference from the ratio of probability
densities at $\lambda=0$ and $\lambda=1$%
.\cite{Abrams_Tuckerman:2006,Wu_Yang:2011} Indeed, our STI\ approach would
also allow obtaining a well converged result by sampling mainly in the regions
of $\lambda$ close to the endpoints $\lambda=0$ and $\lambda=1$ if one used a
modified partition function
\begin{equation}
\tilde{Q}(\lambda)=Q(\lambda)e^{-\beta V_{\mathrm{barr}}(\lambda)},
\end{equation}
where $V_{\mathrm{barr}}(\lambda)$ is a potential that biases the Monte Carlo
chain towards the end points. Running an STI calculation with $J=1$ will lead
to an exact partition function ratio and at the same time use mainly samples from
values of $\lambda$ close to the endpoints. Although such an approach would
avoid the problem of choosing an optimal bin width for the weighted histogram
analysis method (WHAM), an issue discussed in
Ref.~\onlinecite{Kastner_Thiel:2005}, it would, just as WHAM, require
equilibration over the entire $\lambda$ interval $[0,1]$ instead of only over
each subinterval $I_{j}$, which would make it less convenient than the simple
STI presented.

We would also like to mention an alternative
approach allowing to remove the integration error of the isotope
effect entirely, which was proposed recently by Cheng and
Ceriotti\cite{Cheng_Ceriotti:2015} and is a variant of the free
energy perturbation method.\cite{Ceriotti_Markland:2013} Cheng and Ceriotti's
approach employs so-called \textquotedblleft direct
estimators\textquotedblright\ and is particularly suitable for isotope effects
close to unity, which occur frequently, e.g., in the condensed phase, where
only a small fraction of molecules is isotopically substituted. However, for
large isotope effects, such as those discussed here, the direct
estimators tend to have large statistical errors. In the
future,\cite{Karandashev_Vanicek:2017b} we therefore plan to
combine the trick of a stochastic mass change with the direct estimators, in
order to make the latter method practical for large isotope effects as well.

Let us conclude by noting that the
stochastic thermodynamic integration can be combined with
Takahashi-Imada or Suzuki fourth-order
factorizations\cite{Takahashi_Imada:1984,Suzuki:1995,Chin:1997,Jang_Voth:2001}
of the Boltzmann operator, which would allow lowering the path integral
discretization error of the computed isotope effect for a given Trotter number
$P$, and hence a faster convergence to the quantum limit. The combination of
higher-order path integral splittings with standard thermodynamic integration
has been discussed
elsewhere;\cite{Perez_Tuckerman:2011,Buchowiecki_Vanicek:2013,Marsalek_Tuckerman:2014,
Karandashev_Vanicek:2015} as for the extension to stochastic thermodynamic
integration, the main additional change consists in replacing the potential
$V$ in the acceptance probability in Eq.~(\ref{eq:lambda_MC_procedure}) with
an effective potential depending on mass and, in the case of the fourth-order
Suzuki splitting, in an additional factor depending on the imaginary
time-slice index $s$.

\begin{acknowledgments}
We would like to acknowledge the support of this research by the Swiss
National Science Foundation with Grant No. 200020\_150098 and by the EPFL.
Several numerical results were obtained with computational resources provided
by the National Center for Competence in Research \textquotedblleft Molecular
Ultrafast Science and Technology\textquotedblright\ (NCCR MUST).
\end{acknowledgments}

\appendix

\section{\label{app:NMPIMC}Mass-scaled normal modes of the ring polymer and
the derivation of the virial estimator and mass-scaled $\lambda$-move}

In this appendix we outline how transforming from Cartesian
coordinates to mass-scaled normal modes of the ring
polymer\cite{Coalson_Doll:1986,Cao_Berne:1993} leads to simple derivations of
the centroid virial estimator $\left[  dF(\lambda)/d\lambda\right]
_{\text{cv}}$ [Eq.~(\ref{eq:F_cv}) from Subsec.~\ref{subsec:TI}] and the
mass-scaled trial move with respect to the mass parameter $\lambda$
[Eqs.~(\ref{eq:lambda_trial_move})-(\ref{eq:lambda_MC_procedure}) from
Subsec.~\ref{subsec:TI_in_extended_space}].

The mass-scaled normal mode coordinates $\mathbf{a}%
=\{\mathbf{a}^{(1)},\ldots,\mathbf{a}^{(P/2)}\}$ and $\mathbf{b}%
=\{\mathbf{b}^{(1)},\ldots,\mathbf{b}^{(P/2-1)}\}$ can be obtained as
\begin{align}
\mathrm{a}_{i}^{(k)}  &  =\frac{\sqrt{m_{i}}}{P}\sum_{s=1}^{P}\mathrm{r}%
_{i}^{(s)}\cos\left(  \frac{2\pi sk}{P}\right)  ,\,k\in\{1,2,\ldots,P/2\}\\
\mathrm{b}_{i}^{(l)}  &  =\frac{\sqrt{m_{i}}}{P}\sum_{s=1}^{P}\mathrm{r}%
_{i}^{(s)}\sin\left(  \frac{2\pi sl}{P}\right)  ,\,l\in\{1,2,\ldots,P/2-1\},
\end{align}
where $\mathrm{a}_{i}^{(k)}$ and $\mathrm{b}_{i}^{(l)}$ are components of
$\mathbf{a}^{(k)}$ and $\mathbf{b}^{(l)}$ corresponding to particle $i$. This
set of coordinates becomes complete after adding the centroid
$\mathbf{r}^{(C)}=P^{-1}\sum_{s=1}^{P}\mathbf{r}^{(s)}$, which
can also be thought of as the zero-frequency normal mode, and we will refer to
the triple $(\mathbf{a},\mathbf{b},\mathbf{r}^{(C)})$ simply as
$\mathbf{u}$.
Note that, for convenience, we have not mass-scaled $\mathbf{r}^{(C)}$. For
simplicity, we only consider even values of the Trotter number $P$ since the
case of odd $P$ differs in minor details but is otherwise completely analogous.

The original coordinates $\mathbf{r}$ are recovered from the
normal mode coordinates $\mathbf{u}$ via the inverse
transformation
\begin{equation}
\mathrm{r}_{i}^{(s)}=\mathrm{r}_{i}^{(C)}+\frac{1}{\sqrt{m_{i}}}\left\{
(-1)^{s}\mathrm{a}_{i}^{(P/2)}+2\sum_{k=1}^{P/2-1}\left[  \mathrm{a}_{i}%
^{(k)}\cos\left(  \frac{2\pi sk}{P}\right)  +\mathrm{b}_{i}^{(k)}\sin\left(
\frac{2\pi sk}{P}\right)  \right]  \right\}  \label{eq:inverse_transform}%
\end{equation}
with the Jacobian
\begin{equation}
J=\frac{P^{NDP/2}\cdot2^{ND(P/2-1)}}{\left(  \prod_{i=1}^{N}m_{i}\right)
^{D(P-1)/2}}.
\end{equation}
These two expressions can be obtained easily starting from properties of the
real version of the Discrete Fourier
Transform.\cite{Ersoy:1985}

Rewriting the path integral representation of the partition function in terms
of the normal-mode coordinates leads to
\begin{align}
Q_{P}  &  =\int d\mathbf{u}\tilde{\rho}(\mathbf{u})\text{,}\\
\tilde{\rho}  &  =\tilde{C}\exp\left[  -\beta\tilde{\Phi}(\mathbf{u}) \right]
, \label{eq:Qr_PI_rescaled_coord}%
\end{align}
where the new effective potential $\tilde{\Phi}(\mathbf{u})$ and normalization
constant $\tilde{C}$ are given by
\begin{align}
\tilde{\Phi}:=  &  \frac{2P^{2}}{\beta^{2}\hbar^{2}}\left\{  |\mathbf{a}%
^{(P/2)}|^{2}+\sum_{k=1}^{P/2-1}(|\mathbf{a}^{(k)}|^{2}+|\mathbf{b}^{(k)}%
|^{2})\left[  1-\cos\left(  \frac{2\pi k}{P}\right)  \right]  \right\}
\nonumber\\
&  +\frac{1}{P}\sum_{s=1}^{P}V\left[  \mathbf{r}^{(s)}(\mathbf{u}%
,\{m_{i}\})\right]  ,\\
\tilde{C}:=  &  \left(  \frac{P^{2}}{\beta\hbar^{2}\pi}\right)  ^{NDP/2}%
\frac{\left(  \prod_{i=1}^{N}m_{i}\right)  ^{D/2}}{2^{ND}}.
\end{align}
Note that the only term of $\tilde{\Phi}(\mathbf{u})$ depending
on mass is the average of $V(\mathbf{r}^{(s)})$ over the $P$ beads.

With this setup, the centroid virial estimator (\ref{eq:F_cv}) can be obtained
immediately by differentiating the right-hand side of
Eq.~(\ref{eq:Qr_PI_rescaled_coord}) with respect to $\lambda$. To derive the
mass-scaled $\lambda$-move described by
Eqs.~(\ref{eq:lambda_trial_move})-(\ref{eq:lambda_MC_procedure}), we consider
making a trial move with respect to $\lambda$ with $\tilde{\rho
}^{(\lambda)}(\mathbf{u})$ as the probability density while keeping
$\mathbf{u}$ constant. Transforming the corresponding ratio of
probability densities
\begin{equation}%
\begin{split}
\frac{\tilde{\rho}^{(\lambda^{\prime\prime})}(\mathbf{u})}{\tilde{\rho
}^{(\lambda^{\prime})}(\mathbf{u})}=  &  \left[  \prod_{i=1}^{N}\frac
{m_{i}(\lambda^{\prime\prime})}{m_{i}(\lambda^{\prime})}\right]  ^{D/2}\\
&  \times\exp\left(  \frac{\beta}{P}\sum_{s=1}^{P}\{V[\mathbf{r}%
^{(s)}(\mathbf{u},\{m_{i}(\lambda^{\prime})\})]-V[\mathbf{r}^{(s)}%
(\mathbf{u},\{m_{i}(\lambda^{\prime\prime})\})]\}\right)
\end{split}
\end{equation}
back to Cartesian coordinates $\mathbf{r}$ will immediately yield
Eq.~(\ref{eq:lambda_MC_procedure}).

Finally, let us remark that the algorithm used in
Subsec.~\ref{subsec:harmonic} for sampling the harmonic system also uses
normal modes of the ring polymer, albeit not scaled by mass.

\section{\label{app:STI_error_control}Dependence of the error of stochastic
thermodynamic integration on the choice of umbrella biasing potential}

In this appendix we discuss how one may minimize the numerical errors
appearing if the IE is evaluated with STI [via
Eq.~(\ref{eq:lambda_TI_final_formula})] by an appropriate choice of the
umbrella potential. The two errors introduced by the procedure are the
statistical error and integration error due to a finite value of $J$. To
estimate the integration error, we note that $U_{b}(\lambda)$ is independent
of $\mathbf{r}$ and rewrite $\langle\lbrack dF(\lambda)/d\lambda
]_{\mathrm{cv}}\rangle^{I_{j}}$ as
\begin{equation}%
\begin{split}
\left\langle \lbrack dF(\lambda)/d\lambda]_{\mathrm{cv}}\right\rangle ^{I_{j}%
}=  &  \frac{\int_{\lambda_{j-1}}^{\lambda_{j}}d\lambda\int d\mathbf{r}%
\rho^{(\lambda)}(\mathbf{r})[dF(\lambda)/d\lambda]_{\mathrm{cv}}\exp[-\beta
U_{b}(\lambda)]}{\int_{\lambda_{j-1}}^{\lambda_{j}}d\lambda\int d\mathbf{r}%
\rho^{(\lambda)}(\mathbf{r})\exp[-\beta U_{b}(\lambda)]}\\
=  &  \frac{\int_{\lambda_{j-1}}^{\lambda_{j}}d\lambda\exp[-\beta
U_{b}(\lambda)]\int d\mathbf{r}\rho^{(\lambda)}(\mathbf{r})[dF(\lambda
)/d\lambda]_{\mathrm{cv}}}{\int_{\lambda_{j-1}}^{\lambda_{j}}d\lambda
\exp[-\beta U_{b}(\lambda)]\int d\mathbf{r}\rho^{(\lambda)}(\mathbf{r})}\\
=  &  -\frac{\int_{\lambda_{j-1}}^{\lambda_{j}}d\lambda\exp[-\beta
U_{b}(\lambda)+\ln Q(\lambda)]d\ln Q(\lambda)/d\lambda}{\beta\int%
_{\lambda_{j-1}}^{\lambda_{j}}d\lambda\exp[-\beta U_{b}(\lambda)+\ln
Q(\lambda)]}.
\end{split}
\label{eq:F_cv_av_rewritten}%
\end{equation}
Now let us consider several possible choices for the umbrella potential; an
impatient reader should skip the subsection on a piecewise constant umbrella
potential since we show that the most useful in practice is the
\emph{piecewise linear} umbrella potential.

\subsection{Exact umbrella potential}

Suppose that one can find the ideal, \textquotedblleft exact\textquotedblright%
\ umbrella potential
\begin{equation}
U_{b,\mathrm{exact}}(\lambda):=\beta^{-1}\ln Q(\lambda).
\label{eq:def_Ub_exact}%
\end{equation}
Using this exact umbrella potential amounts to the substitution $U_{b}%
(\lambda)=U_{b,\mathrm{exact}}(\lambda)$ in Eq.~(\ref{eq:F_cv_av_rewritten})
and gives
\begin{equation}
\left\langle \lbrack dF(\lambda)/d\lambda]_{\mathrm{cv}}\right\rangle ^{I_{j}%
}=-\frac{\ln Q(\lambda_{j})-\ln Q(\lambda_{j-1})}{\beta(\lambda_{j}%
-\lambda_{j-1})}. \label{eq:Delta_F_exact_Ub}%
\end{equation}
Since $\lambda_{j}-\lambda_{j-1}=J^{-1}$, in this ideal situation
Eq.~(\ref{eq:lambda_TI_final_formula}) will yield the exact partition function
ratio at any value of $J$.

\subsection{Piecewise constant umbrella potential}

Unfortunately, in a realistic calculation this ideal potential
$U_{b,\text{exact}}(\lambda)$ is not available and one must make do with an
approximation. The simplest choice is a piecewise constant potential%
\begin{equation}
U_{b,\mathrm{p.const.}}(\lambda):=U_{b,j}\text{ for }\lambda\in(\lambda
_{j},\lambda_{j-1}). \label{eq:def_Ub_piecewise_constant}%
\end{equation}
To simplify the following algebra we introduce a symbol
\begin{equation}
\Delta(\lambda):=\ln Q(\lambda)-\ln Q(\overline{\lambda}_{j})
\label{eq:Delta_lnQ_defined}%
\end{equation}
and note that after the substitution $U_{b}(\lambda)=U_{b,\mathrm{p.const.}%
}(\lambda)$ the constant factor $\exp[-\beta U_{b,j}+\ln Q(\overline{\lambda
}_{j})]$ will cancel out between the numerator and denominator of
Eq.~(\ref{eq:F_cv_av_rewritten}), leading to a simplified expression
\begin{equation}
\left\langle [dF(\lambda)/d\lambda]_{\mathrm{cv}}\right\rangle ^{I_{j}}=-\frac
{\int_{\lambda_{j-1}}^{\lambda_{j}}d\lambda\exp[\Delta(\lambda)]d\ln
Q(\lambda)/d\lambda}{\beta\int_{\lambda_{j-1}}^{\lambda_{j}}d\lambda
\exp[\Delta(\lambda)]}. \label{eq:Delta_F_piecewise_const}%
\end{equation}
Although it was not important for the derivation of the last equation, it is
worthwhile to mention that the constants $U_{b,j}$ are determined in the
simulation from the equation%
\begin{equation}
U_{b,j+1}=U_{b,j}+\frac{\left\langle [dF(\lambda)/d\lambda]_{\mathrm{cv}}\right\rangle
^{I_{j}}+\left\langle [dF(\lambda)/d\lambda]_{\mathrm{cv}}\right\rangle ^{I_{j+1}}}%
{2J}. \label{eq:Ubj}%
\end{equation}
Upon changing variables from $\lambda$ to $\Delta(\lambda)$, the numerator of
Eq.~(\ref{eq:Delta_F_piecewise_const}) becomes
\begin{equation}
\int_{\Delta(\lambda_{j-1})}^{\Delta(\lambda_{j})}e^{\Delta(\lambda)}%
d\Delta(\lambda)=e^{\Delta(\lambda_{j})}-e^{\Delta(\lambda_{j-1})},
\end{equation}
hence
\begin{equation}
\left\langle \lbrack dF(\lambda)/d\lambda]_{\mathrm{cv}}\right\rangle ^{I_{j}%
}=-\frac{f(\lambda_{j})-f(\lambda_{j-1})}{\beta\int_{\lambda_{j-1}}%
^{\lambda_{j}}[1+f(\lambda)]d\lambda},
\label{eq:Delta_F_piecewise_const_final}%
\end{equation}
where we defined a function
\begin{equation}
f(\lambda):=e^{\Delta(\lambda)}-1, \label{eq:f_definition}%
\end{equation}
whose Taylor series expansion about $\overline{\lambda}_{j}$,
\begin{equation}
f(\lambda)=f^{\prime}(\overline{\lambda}_{j})(\lambda-\overline{\lambda}%
_{j})+\frac{f^{\prime\prime}(\overline{\lambda}_{j})}{2}(\lambda
-\overline{\lambda}_{j})^{2}+\frac{f^{\prime\prime\prime}(\overline{\lambda
}_{j})}{6}(\lambda-\overline{\lambda}_{j})^{3}+\mathcal{O}[(\lambda
-\overline{\lambda}_{j})^{4}], \label{eq:f_Taylor_series}%
\end{equation}
will be used in the following. To see how good an approximation the piecewise
constant potential gives, let us compare the numerators and denominators of
Eqs.~(\ref{eq:Delta_F_exact_Ub}) and (\ref{eq:Delta_F_piecewise_const_final}).
The difference of the denominators is
\begin{equation}
\beta\int_{\lambda_{j-1}}^{\lambda_{j}}[1+f(\lambda)]d\lambda-\beta
(\lambda_{j}-\lambda_{j-1})=\frac{\beta f^{\prime\prime}(\overline{\lambda
}_{j})}{24J^{3}}+\mathcal{O}(J^{-4}). \label{eq:denominator_error}%
\end{equation}
Noting that $f(\lambda)=\mathcal{O}(\lambda-\overline{\lambda}_{j})$ and
Taylor expanding the logarithm, we find the difference of the numerators to
be
\begin{equation}%
\begin{split}
&  f(\lambda_{j})-f(\lambda_{j-1})-\ln[1+f(\lambda_{j})]+\ln[1+f(\lambda
_{j-1})]\\
=  &  \frac{f(\lambda_{j})^{2}-f(\lambda_{j-1})^{2}}{2}-\frac{f(\lambda
_{j})^{3}-f(\lambda_{j-1})^{3}}{3}+\mathcal{O}(J^{-4})\\
=  &  \frac{[f(\lambda_{j})-f(\lambda_{j-1})][f(\lambda_{j})+f(\lambda
_{j-1})]}{2}\\
&  -\frac{f^{\prime}(\overline{\lambda}_{j})^{3}(\lambda_{j}-\lambda
_{j-1})^{3}}{12}+\mathcal{O}(J^{-4})\\
=  &  \frac{f^{\prime}(\overline{\lambda}_{j})f^{\prime\prime}(\overline
{\lambda}_{j})}{8J^{3}}-\frac{f^{\prime}(\overline{\lambda}_{j})^{3}}{12J^{3}}+\mathcal{O}(J^{-4}).
\end{split}
\label{eq:numerator_error}%
\end{equation}
Since both the numerator and denominator of Eq.~(\ref{eq:Delta_F_exact_Ub})
are $\mathcal{O}(J^{-1})$, and since the errors in
Eq.~(\ref{eq:Delta_F_piecewise_const_final}) of both the denominator
[Eq.~(\ref{eq:denominator_error})] and numerator
[Eq.~(\ref{eq:numerator_error})] are $\mathcal{O}(J^{-3})$, the overall error
is $\mathcal{O}(J^{-2})$, that is, for an umbrella potential constant over
each $I_{j}$
\begin{equation}
\langle\lbrack dF(\lambda)/d\lambda]_{\mathrm{cv}}\rangle^{I_{j}}=-\frac{\ln
Q(\lambda_{j})-\ln Q(\lambda_{j-1})}{\beta(\lambda_{j}-\lambda_{j-1}%
)}+\mathcal{O}(J^{-2}). \label{eq:lambda_TI_midpoint_term_error}%
\end{equation}
In conclusion, for the piecewise constant biasing potential
Eq.~(\ref{eq:lambda_TI_final_formula}) will have an error $\mathcal{O}%
(J^{-2})$:
\begin{equation}
\exp\left\{  -\frac{\beta}{J}\sum_{j=1}^{J}\left\langle \left[  dF(\lambda
)/d\lambda\right]  _{\text{cv}}\right\rangle ^{I_{j}}\right\}  =\frac
{Q_{P}^{(B)}}{Q_{P}^{(A)}}+\mathcal{O}(J^{-2}).
\label{eq:lambda_TI_final_formula_midpoint}%
\end{equation}
As discussed in Subsec.~(\ref{subsec:TI_in_extended_space}), it is easy to use
really large values of $J$ during the calculation, therefore an $\mathcal{O}%
(J^{-2})$ error is not an issue. Yet, it is still worthwhile to try to
optimize the procedure in order to go beyond an $\mathcal{O}(J^{-2})$ error.

\subsection{Piecewise linear umbrella potential}

The obvious \textquotedblleft first\textquotedblright\ improvement is
introducing a piecewise linear potential. A remarkable fact about the
resulting procedure is that it yields an exactly zero integration error, and
this is true to all orders in $J$. Indeed, if we introduce a $U_{b}%
(\lambda)=U_{b,\mathrm{p.lin.}}(\lambda)$, where
\begin{equation}
U_{b,\mathrm{p.lin.}}(\lambda):=U_{b,j}-\langle\lbrack dF(\lambda
)/d\lambda]_{\mathrm{cv}}\rangle^{I_{j}}(\lambda-\overline{\lambda}_{j}),
\label{eq:U_b_current}%
\end{equation}
then the constant factor $\exp(-\beta\{U_{b,j}-\langle\lbrack dF(\lambda
)/d\lambda]_{\mathrm{cv}}\rangle^{I_{j}}\overline{\lambda}_{j}\})$ will cancel
between the numerator and denominator of Eq.~(\ref{eq:F_cv_av_rewritten}),
giving
\begin{equation}
\left\langle [dF(\lambda)/d\lambda]_{\mathrm{cv}}\right\rangle ^{I_{j}}=-\frac
{\int_{\lambda_{j-1}}^{\lambda_{j}}d\lambda\exp\{\beta\langle\lbrack
dF(\lambda)/d\lambda]_{\mathrm{cv}}\rangle^{I_{j}}\lambda+\ln Q(\lambda)\}d\ln
Q(\lambda)/d\lambda}{\beta\int_{\lambda_{j-1}}^{\lambda_{j}}d\lambda
\exp\{\beta\langle\lbrack dF(\lambda)/d\lambda]_{\mathrm{cv}}\rangle^{I_{j}%
}\lambda+\ln Q(\lambda)\}}.
\end{equation}
Multiplying both sides of the equation by the denominator and rearranging
yields an identity%
\begin{align}
0  &  =\int_{\lambda_{j-1}}^{\lambda_{j}}\exp\{\beta\langle\lbrack
dF(\lambda)/d\lambda]_{\mathrm{cv}}\rangle^{I_{j}}\lambda+\ln Q(\lambda
)\}\{\beta\left\langle \lbrack dF(\lambda)/d\lambda]_{\mathrm{cv}%
}\right\rangle ^{I_{j}}d\lambda+d\ln Q(\lambda)\}\\
&  =\int_{g(\lambda_{j-1})}^{g(\lambda_{j})}dg(\lambda)e^{g(\lambda
)}=e^{g(\lambda_{j})}-e^{g(\lambda_{j-1})},
\end{align}
where we have introduced a function $g(\lambda):=\beta\langle\lbrack
dF(\lambda)/d\lambda]_{\mathrm{cv}}\rangle^{I_{j}}\lambda+\ln Q(\lambda)$. The
last equality means $g(\lambda_{j})=g(\lambda_{j-1})$, leading to
\begin{equation}
\langle\lbrack dF(\lambda)/d\lambda]_{\mathrm{cv}}\rangle^{I_{j}}=-\frac{\ln
Q(\lambda_{j})-\ln Q(\lambda_{j-1})}{\beta(\lambda_{j}-\lambda_{j-1}%
)},\label{eq:final_F_av_rewritten}%
\end{equation}
which is, remarkably, the same as Eq.~(\ref{eq:Delta_F_exact_Ub}) for the
\emph{exact} umbrella potential.

Of course, the definition of the piecewise linear umbrella potential in
Eq.~(\ref{eq:U_b_current}) is recursive, and therefore can only be evaluated
by an iterative algorithm, but this should not cause a great concern since any
biasing potential $U_{b}(\lambda)$, regardless of its type, cannot be known
\textit{a priori} (in particular, even the piecewise constant umbrella
potential must be constructed iteratively).

As already mentioned in Subsec.~(\ref{subsec:TI_in_extended_space}), making
$U_{b}(\lambda)$ not only piecewise linear [by satisfying
Eq.~(\ref{eq:U_b_current})] but also continuous [by choosing the constants
$U_{b,j}$ from Eq.~(\ref{eq:Ubj})] allows to approach an optimal statistical
error. An
analytical analysis of the statistical error is more involved; instead, in the
following subsection we show numerically that the statistical error is
approximately independent of the choice of the umbrella potential and converges to 
a limit as $ J$ is increased---in
particular, the piecewise linear umbrella potential permits reducing the
integration to zero without increasing the statistical error.

\subsection{Numerical tests}

As the piecewise linear umbrella potential $U_{b,\mathrm{p.lin.}}(\lambda)$
defined by Eq.~(\ref{eq:U_b_choice}) yields a zero integration error and can
be obtained iteratively in any system, it was this potential that was used in
the production runs in the rest of the paper. To clearly demonstrate the advantages of
$U_{b,\mathrm{p.lin.}}(\lambda)$, in this subsection we compare the different choices of the
umbrella potential on the harmonic system from Subsec.~\ref{subsec:harmonic},
for which which even the exact umbrella potential (\ref{eq:def_Ub_exact})\ is
available since $Q(\lambda)$ is known analytically.

\begin{figure}
[tbp]%
\centering\includegraphics[width=\textwidth]{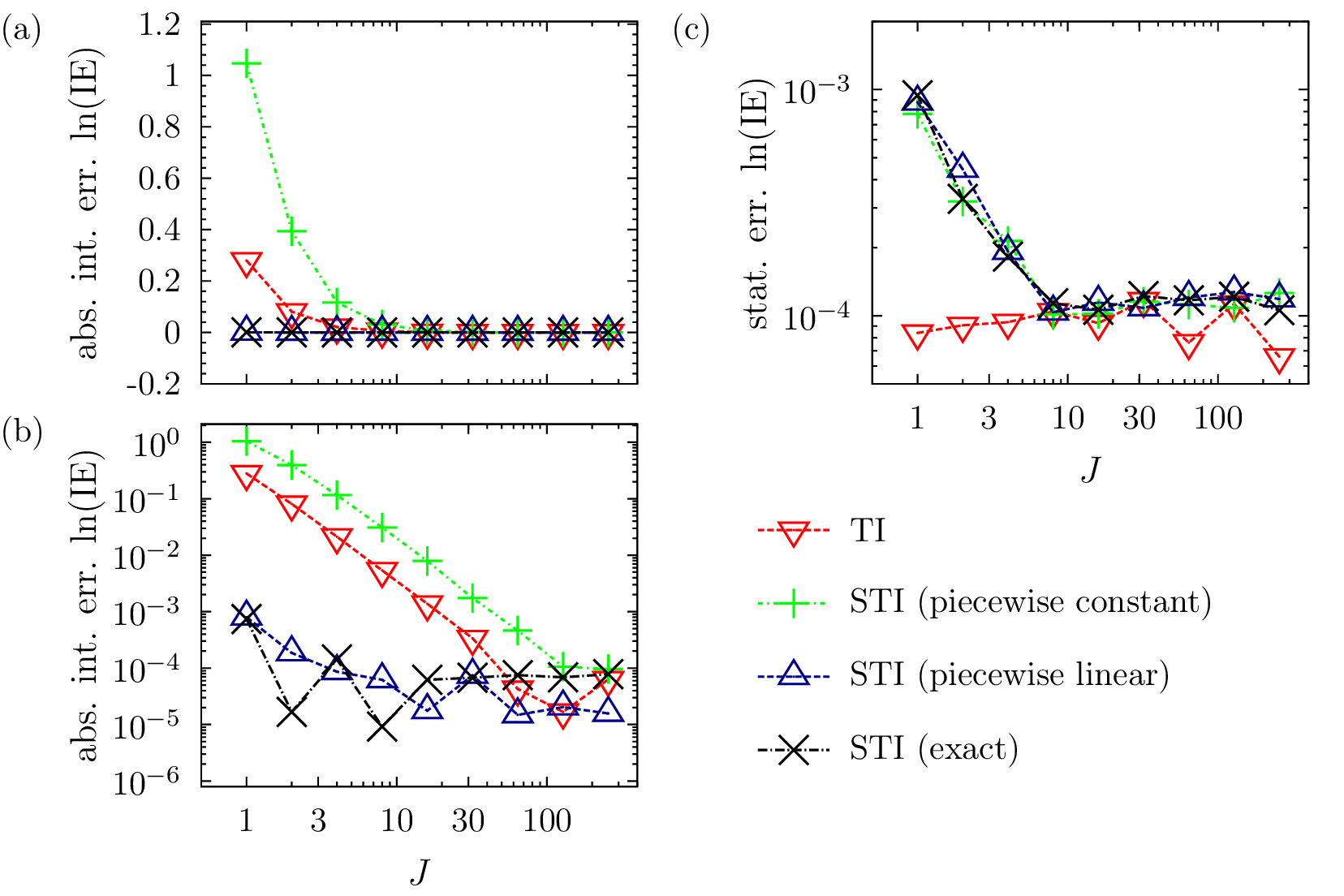}
\caption{Impact of the umbrella biasing potential
on the numerical errors of a simulation. The figure shows the dependence of
integration errors [(a) with linear, (b) with logarithmic scale]
and statistical errors [panel (c)] on the number $ J$ of $ \lambda$
intervals used in the
standard thermodynamic integration (TI)
and stochastic thermodynamic integration (STI)
employing various types of umbrella potentials
mentioned in parentheses: ``piecewise constant'' uses an
umbrella potential (\ref{eq:def_Ub_piecewise_constant}) constant over each of the $ J$ intervals,
``piecewise linear'' potential is given by Eqs.~(\ref{eq:U_b_choice}) or (\ref{eq:U_b_current}), and ``exact''
corresponds to the unrealistic situation when one knows the exact, ideal umbrella potential
(\ref{eq:def_Ub_exact}).
} \label{fig:sho_diff_STI_proc}
\end{figure}

The results are presented in Fig.~\ref{fig:sho_diff_STI_proc}, which shows the
dependence of integration and statistical errors on $J$. [Note that all
methods employed the linear interpolation of mass given by
Eq.~(\ref{eq:m_linear_interpol}) and that the values obtained with TI\ and
with STI with a piecewise linear potential\ are the same as those already
presented in Subsec.~\ref{subsec:harmonic}.]

As predicted above, the integration error of STI\ appears to be zero for both
the ideal, exact\ umbrella potential (\ref{eq:def_Ub_exact}) and for the
piecewise linear umbrella potential (\ref{eq:U_b_current}) [panels (a) and
(b)], whereas both TI with the midpoint rule and STI with the piecewise
constant umbrella potential (\ref{eq:def_Ub_piecewise_constant}) exhibit an
$\mathcal{O}(J^{-2})$ integration error [panel (b)]. Note that for a given $J$
the integration error of STI\ with the piecewise constant potential is even
larger than the error of TI using the midpoint rule; however, this does not
imply that STI\ is less efficient than TI since in realistic calculations STI
can be used with much larger values of $J$ than TI, without increasing the
statistical error or the computational cost. Finally, note that the different
choices of the umbrella potential do not significantly affect the statistical
error [see panel (c)].

\color{BLACK}

%\bibliographystyle{aipnum4-1}
%\bibliography{EIE_extended_space}

\begin{thebibliography}{62}%
\makeatletter
\providecommand \@ifxundefined [1]{%
 \@ifx{#1\undefined}
}%
\providecommand \@ifnum [1]{%
 \ifnum #1\expandafter \@firstoftwo
 \else \expandafter \@secondoftwo
 \fi
}%
\providecommand \@ifx [1]{%
 \ifx #1\expandafter \@firstoftwo
 \else \expandafter \@secondoftwo
 \fi
}%
\providecommand \natexlab [1]{#1}%
\providecommand \enquote  [1]{``#1''}%
\providecommand \bibnamefont  [1]{#1}%
\providecommand \bibfnamefont [1]{#1}%
\providecommand \citenamefont [1]{#1}%
\providecommand \href@noop [0]{\@secondoftwo}%
\providecommand \href [0]{\begingroup \@sanitize@url \@href}%
\providecommand \@href[1]{\@@startlink{#1}\@@href}%
\providecommand \@@href[1]{\endgroup#1\@@endlink}%
\providecommand \@sanitize@url [0]{\catcode `\\12\catcode `\$12\catcode
  `\&12\catcode `\#12\catcode `\^12\catcode `\_12\catcode `\%12\relax}%
\providecommand \@@startlink[1]{}%
\providecommand \@@endlink[0]{}%
\providecommand \url  [0]{\begingroup\@sanitize@url \@url }%
\providecommand \@url [1]{\endgroup\@href {#1}{\urlprefix }}%
\providecommand \urlprefix  [0]{URL }%
\providecommand \Eprint [0]{\href }%
\providecommand \doibase [0]{http://dx.doi.org/}%
\providecommand \selectlanguage [0]{\@gobble}%
\providecommand \bibinfo  [0]{\@secondoftwo}%
\providecommand \bibfield  [0]{\@secondoftwo}%
\providecommand \translation [1]{[#1]}%
\providecommand \BibitemOpen [0]{}%
\providecommand \bibitemStop [0]{}%
\providecommand \bibitemNoStop [0]{.\EOS\space}%
\providecommand \EOS [0]{\spacefactor3000\relax}%
\providecommand \BibitemShut  [1]{\csname bibitem#1\endcsname}%
\let\auto@bib@innerbib\@empty
%</preamble>
\bibitem [{\citenamefont {Wolfsberg}\ \emph {et~al.}(2010)\citenamefont
  {Wolfsberg}, \citenamefont {Hook}, \citenamefont {Paneth},\ and\
  \citenamefont {Rebelo}}]{Wolfsberg_Rebelo:2010}%
  \BibitemOpen
  \bibfield  {author} {\bibinfo {author} {\bibfnamefont {M.}~\bibnamefont
  {Wolfsberg}}, \bibinfo {author} {\bibfnamefont {W.~A.~V.}\ \bibnamefont
  {Hook}}, \bibinfo {author} {\bibfnamefont {P.}~\bibnamefont {Paneth}}, \ and\
  \bibinfo {author} {\bibfnamefont {L.~P.~N.}\ \bibnamefont {Rebelo}},\
  }\href@noop {} {\emph {\bibinfo {title} {Isotope Effects in the Chemical,
  Geological and Bio Sciences}}}\ (\bibinfo  {publisher} {McGraw-Hill},\
  \bibinfo {year} {2010})\BibitemShut {NoStop}%
\bibitem [{\citenamefont {Urey}(1947)}]{Urey:1947}%
  \BibitemOpen
  \bibfield  {author} {\bibinfo {author} {\bibfnamefont {H.~C.}\ \bibnamefont
  {Urey}},\ }\href
  {http://pubs.rsc.org/en/content/articlepdf/1947/jr/jr9470000562} {\bibfield
  {journal} {\bibinfo  {journal} {J. Chem. Soc.}\ }\textbf {\bibinfo {volume}
  {1947}},\ \bibinfo {eid} {562} (\bibinfo {year} {1947})}\BibitemShut
  {NoStop}%
\bibitem [{\citenamefont {McKenzie}, \citenamefont {Athokpam},\ and\
  \citenamefont {Ramesh}(2015)}]{McKenzie_Ramesh:2015}%
  \BibitemOpen
  \bibfield  {author} {\bibinfo {author} {\bibfnamefont {R.~M.}\ \bibnamefont
  {McKenzie}}, \bibinfo {author} {\bibfnamefont {B.}~\bibnamefont {Athokpam}},
  \ and\ \bibinfo {author} {\bibfnamefont {S.~G.}\ \bibnamefont {Ramesh}},\
  }\href {http://dx.doi.org/10.1063/1.4927391} {\bibfield  {journal} {\bibinfo
  {journal} {J. Chem. Phys.}\ }\textbf {\bibinfo {volume} {143}},\ \bibinfo
  {eid} {044309} (\bibinfo {year} {2015})}\BibitemShut {NoStop}%
\bibitem [{\citenamefont {Janak}\ and\ \citenamefont
  {Parkin}(2003)}]{Janak_Parkin:2003}%
  \BibitemOpen
  \bibfield  {author} {\bibinfo {author} {\bibfnamefont {K.~E.}\ \bibnamefont
  {Janak}}\ and\ \bibinfo {author} {\bibfnamefont {G.}~\bibnamefont {Parkin}},\
  }\href {http://pubs.acs.org/doi/abs/10.1021/ja0362611} {\bibfield  {journal}
  {\bibinfo  {journal} {J. Am. Chem. Soc.}\ }\textbf {\bibinfo {volume}
  {125}},\ \bibinfo {eid} {13219} (\bibinfo {year} {2003})}\BibitemShut
  {NoStop}%
\bibitem [{\citenamefont {Webb}\ and\ \citenamefont {{Miller
  III}}(2014)}]{Webb_Miller:2014}%
  \BibitemOpen
  \bibfield  {author} {\bibinfo {author} {\bibfnamefont {M.~A.}\ \bibnamefont
  {Webb}}\ and\ \bibinfo {author} {\bibfnamefont {T.~F.}\ \bibnamefont {{Miller
  III}}},\ }\href@noop {} {\bibfield  {journal} {\bibinfo  {journal} {J. Phys.
  Chem. A}\ }\textbf {\bibinfo {volume} {118}},\ \bibinfo {eid} {467} (\bibinfo
  {year} {2014})}\BibitemShut {NoStop}%
\bibitem [{\citenamefont {Richet}, \citenamefont {Bottinga},\ and\
  \citenamefont {Javoy}(1977)}]{Richet_Javoy:1977}%
  \BibitemOpen
  \bibfield  {author} {\bibinfo {author} {\bibfnamefont {P.}~\bibnamefont
  {Richet}}, \bibinfo {author} {\bibfnamefont {Y.}~\bibnamefont {Bottinga}}, \
  and\ \bibinfo {author} {\bibfnamefont {M.}~\bibnamefont {Javoy}},\
  }\href@noop {} {\bibfield  {journal} {\bibinfo  {journal} {Ann. Rev. Earth
  Planet. Sci.}\ }\textbf {\bibinfo {volume} {5}},\ \bibinfo {eid} {65}
  (\bibinfo {year} {1977})}\BibitemShut {NoStop}%
\bibitem [{\citenamefont {Barone}(2004)}]{Barone:2004}%
  \BibitemOpen
  \bibfield  {author} {\bibinfo {author} {\bibfnamefont {V.}~\bibnamefont
  {Barone}},\ }\href@noop {} {\bibfield  {journal} {\bibinfo  {journal} {J.
  Chem. Phys.}\ }\textbf {\bibinfo {volume} {120}},\ \bibinfo {eid} {3059}
  (\bibinfo {year} {2004})}\BibitemShut {NoStop}%
\bibitem [{\citenamefont {Liu}, \citenamefont {Tossell},\ and\ \citenamefont
  {Liu}(2010)}]{Liu_Liu:2010}%
  \BibitemOpen
  \bibfield  {author} {\bibinfo {author} {\bibfnamefont {Q.}~\bibnamefont
  {Liu}}, \bibinfo {author} {\bibfnamefont {J.~A.}\ \bibnamefont {Tossell}}, \
  and\ \bibinfo {author} {\bibfnamefont {Y.}~\bibnamefont {Liu}},\ }\href@noop
  {} {\bibfield  {journal} {\bibinfo  {journal} {Geochim. Cosmochim. Acta}\
  }\textbf {\bibinfo {volume} {74}},\ \bibinfo {eid} {6965} (\bibinfo {year}
  {2010})}\BibitemShut {NoStop}%
\bibitem [{\citenamefont {Feynman}\ and\ \citenamefont
  {Hibbs}(1965)}]{Feynman_Hibbs:1965}%
  \BibitemOpen
  \bibfield  {author} {\bibinfo {author} {\bibfnamefont {R.~P.}\ \bibnamefont
  {Feynman}}\ and\ \bibinfo {author} {\bibfnamefont {A.~R.}\ \bibnamefont
  {Hibbs}},\ }\href@noop {} {\emph {\bibinfo {title} {Quantum mechanics and
  path integrals}}}\ (\bibinfo  {publisher} {McGraw-Hill},\ \bibinfo {year}
  {1965})\BibitemShut {NoStop}%
\bibitem [{\citenamefont {Chandler}\ and\ \citenamefont
  {Wolynes}(1981)}]{Chandler_Wolynes:1981}%
  \BibitemOpen
  \bibfield  {author} {\bibinfo {author} {\bibfnamefont {D.}~\bibnamefont
  {Chandler}}\ and\ \bibinfo {author} {\bibfnamefont {P.~G.}\ \bibnamefont
  {Wolynes}},\ }\href {http://dx.doi.org/10.1063/1.441588} {\bibfield
  {journal} {\bibinfo  {journal} {J. Chem. Phys.}\ }\textbf {\bibinfo {volume}
  {74}},\ \bibinfo {pages} {4078} (\bibinfo {year} {1981})}\BibitemShut
  {NoStop}%
\bibitem [{\citenamefont {Ceperley}(1995)}]{Ceperley:1995}%
  \BibitemOpen
  \bibfield  {author} {\bibinfo {author} {\bibfnamefont {D.~M.}\ \bibnamefont
  {Ceperley}},\ }\href
  {http://journals.aps.org/rmp/pdf/10.1103/RevModPhys.67.279} {\bibfield
  {journal} {\bibinfo  {journal} {Rev. Mod. Phys.}\ }\textbf {\bibinfo {volume}
  {67}},\ \bibinfo {eid} {279} (\bibinfo {year} {1995})}\BibitemShut {NoStop}%
\bibitem [{\citenamefont {Kirkwood}(1935)}]{Kirkwood:1935}%
  \BibitemOpen
  \bibfield  {author} {\bibinfo {author} {\bibfnamefont {J.~G.}\ \bibnamefont
  {Kirkwood}},\ }\href {http://dx.doi.org/10.1063/1.1749657} {\bibfield
  {journal} {\bibinfo  {journal} {J. Chem. Phys.}\ }\textbf {\bibinfo {volume}
  {3}},\ \bibinfo {eid} {300} (\bibinfo {year} {1935})}\BibitemShut {NoStop}%
\bibitem [{\citenamefont {Frenkel}\ and\ \citenamefont
  {Smit}(2002)}]{Frenkel_Smit:2002}%
  \BibitemOpen
  \bibfield  {author} {\bibinfo {author} {\bibfnamefont {D.}~\bibnamefont
  {Frenkel}}\ and\ \bibinfo {author} {\bibfnamefont {B.}~\bibnamefont {Smit}},\
  }\href@noop {} {\emph {\bibinfo {title} {Understanding Molecular
  Simulation}}}\ (\bibinfo  {publisher} {Academic Press},\ \bibinfo {year}
  {2002})\BibitemShut {NoStop}%
\bibitem [{\citenamefont {Tuckerman}(2010)}]{Tuckerman:2010}%
  \BibitemOpen
  \bibfield  {author} {\bibinfo {author} {\bibfnamefont {M.~E.}\ \bibnamefont
  {Tuckerman}},\ }\href@noop {} {\emph {\bibinfo {title} {Statistical
  Mechanics: Theory and Molecular Simulation}}}\ (\bibinfo  {publisher} {Oxford
  University Press},\ \bibinfo {year} {2010})\BibitemShut {NoStop}%
\bibitem [{\citenamefont {Van\'{i}\v{c}ek}\ \emph {et~al.}(2005)\citenamefont
  {Van\'{i}\v{c}ek}, \citenamefont {Miller}, \citenamefont {Castillo},\ and\
  \citenamefont {Aoiz}}]{Vanicek_Aoiz:2005}%
  \BibitemOpen
  \bibfield  {author} {\bibinfo {author} {\bibfnamefont {J.}~\bibnamefont
  {Van\'{i}\v{c}ek}}, \bibinfo {author} {\bibfnamefont {W.~H.}\ \bibnamefont
  {Miller}}, \bibinfo {author} {\bibfnamefont {J.~F.}\ \bibnamefont
  {Castillo}}, \ and\ \bibinfo {author} {\bibfnamefont {F.~J.}\ \bibnamefont
  {Aoiz}},\ }\href {http://dx.doi.org/10.1063/1.1946740} {\bibfield  {journal}
  {\bibinfo  {journal} {J. Chem. Phys.}\ }\textbf {\bibinfo {volume} {123}},\
  \bibinfo {eid} {054108} (\bibinfo {year} {2005})}\BibitemShut {NoStop}%
\bibitem [{\citenamefont {Van\'{i}\v{c}ek}\ and\ \citenamefont
  {Miller}(2007)}]{Vanicek_Miller:2007}%
  \BibitemOpen
  \bibfield  {author} {\bibinfo {author} {\bibfnamefont {J.}~\bibnamefont
  {Van\'{i}\v{c}ek}}\ and\ \bibinfo {author} {\bibfnamefont {W.~H.}\
  \bibnamefont {Miller}},\ }\href {http://dx.doi.org/10.1063/1.2768930}
  {\bibfield  {journal} {\bibinfo  {journal} {J. Chem. Phys.}\ }\textbf
  {\bibinfo {volume} {127}},\ \bibinfo {eid} {114309} (\bibinfo {year}
  {2007})}\BibitemShut {NoStop}%
\bibitem [{\citenamefont {Zimmermann}\ and\ \citenamefont
  {Van\'{i}\v{c}ek}(2009)}]{Zimmermann_Vanicek:2009}%
  \BibitemOpen
  \bibfield  {author} {\bibinfo {author} {\bibfnamefont {T.}~\bibnamefont
  {Zimmermann}}\ and\ \bibinfo {author} {\bibfnamefont {J.}~\bibnamefont
  {Van\'{i}\v{c}ek}},\ }\href
  {http://scitation.aip.org/content/aip/journal/jcp/131/2/10.1063/1.3167353}
  {\bibfield  {journal} {\bibinfo  {journal} {J. Chem. Phys.}\ }\textbf
  {\bibinfo {volume} {131}},\ \bibinfo {eid} {024111} (\bibinfo {year}
  {2009})}\BibitemShut {NoStop}%
\bibitem [{\citenamefont {Zimmermann}\ and\ \citenamefont
  {Van\'{i}\v{c}ek}(2010)}]{Zimmermann_Vanicek:2010}%
  \BibitemOpen
  \bibfield  {author} {\bibinfo {author} {\bibfnamefont {T.}~\bibnamefont
  {Zimmermann}}\ and\ \bibinfo {author} {\bibfnamefont {J.}~\bibnamefont
  {Van\'{i}\v{c}ek}},\ }\href
  {http://link.springer.com/article/10.1007/s00894-010-0711-y} {\bibfield
  {journal} {\bibinfo  {journal} {J. Mol. Model.}\ }\textbf {\bibinfo {volume}
  {16}},\ \bibinfo {eid} {1779} (\bibinfo {year} {2010})}\BibitemShut {NoStop}%
\bibitem [{\citenamefont {P\'{e}rez}\ and\ \citenamefont {von
  Lilienfeld}(2011)}]{Perez_Lilienfeld:2011}%
  \BibitemOpen
  \bibfield  {author} {\bibinfo {author} {\bibfnamefont {A.}~\bibnamefont
  {P\'{e}rez}}\ and\ \bibinfo {author} {\bibfnamefont {O.~A.}\ \bibnamefont
  {von Lilienfeld}},\ }\href {http://dx.doi.org/10.1021/ct2000556} {\bibfield
  {journal} {\bibinfo  {journal} {J. Chem. Theory Comput.}\ }\textbf {\bibinfo
  {volume} {7}},\ \bibinfo {pages} {2358} (\bibinfo {year} {2011})}\BibitemShut
  {NoStop}%
\bibitem [{\citenamefont {Ceriotti}\ and\ \citenamefont
  {Markland}(2013)}]{Ceriotti_Markland:2013}%
  \BibitemOpen
  \bibfield  {author} {\bibinfo {author} {\bibfnamefont {M.}~\bibnamefont
  {Ceriotti}}\ and\ \bibinfo {author} {\bibfnamefont {T.~E.}\ \bibnamefont
  {Markland}},\ }\href {http://link.aip.org/link/?JCP/138/014112/1} {\bibfield
  {journal} {\bibinfo  {journal} {J. Chem. Phys.}\ }\textbf {\bibinfo {volume}
  {138}},\ \bibinfo {eid} {014112} (\bibinfo {year} {2013})}\BibitemShut
  {NoStop}%
\bibitem [{\citenamefont {Mar\v{s}\'{a}lek}\ \emph {et~al.}(2014)\citenamefont
  {Mar\v{s}\'{a}lek}, \citenamefont {Chen}, \citenamefont {Dupuis},
  \citenamefont {Benoit}, \citenamefont {M\'{e}heut}, \citenamefont
  {Ba\v{c}i\'{c}},\ and\ \citenamefont {Tuckerman}}]{Marsalek_Tuckerman:2014}%
  \BibitemOpen
  \bibfield  {author} {\bibinfo {author} {\bibfnamefont {O.}~\bibnamefont
  {Mar\v{s}\'{a}lek}}, \bibinfo {author} {\bibfnamefont {P.-Y.}\ \bibnamefont
  {Chen}}, \bibinfo {author} {\bibfnamefont {R.}~\bibnamefont {Dupuis}},
  \bibinfo {author} {\bibfnamefont {M.}~\bibnamefont {Benoit}}, \bibinfo
  {author} {\bibfnamefont {M.}~\bibnamefont {M\'{e}heut}}, \bibinfo {author}
  {\bibfnamefont {Z.}~\bibnamefont {Ba\v{c}i\'{c}}}, \ and\ \bibinfo {author}
  {\bibfnamefont {M.~E.}\ \bibnamefont {Tuckerman}},\ }\href
  {http://dx.doi.org/10.1021/ct400911m} {\bibfield  {journal} {\bibinfo
  {journal} {J. Chem. Theory Comput.}\ }\textbf {\bibinfo {volume} {10}},\
  \bibinfo {eid} {1440} (\bibinfo {year} {2014})}\BibitemShut {NoStop}%
\bibitem [{\citenamefont {Liu}\ and\ \citenamefont
  {Berne}(1993)}]{Liu_Berne:1993}%
  \BibitemOpen
  \bibfield  {author} {\bibinfo {author} {\bibfnamefont {Z.}~\bibnamefont
  {Liu}}\ and\ \bibinfo {author} {\bibfnamefont {B.~J.}\ \bibnamefont
  {Berne}},\ }\href {http://dx.doi.org/10.1063/1.465904} {\bibfield  {journal}
  {\bibinfo  {journal} {J. Chem. Phys.}\ }\textbf {\bibinfo {volume} {99}},\
  \bibinfo {eid} {6071} (\bibinfo {year} {1993})}\BibitemShut {NoStop}%
\bibitem [{\citenamefont {Kong}\ and\ \citenamefont {{Brooks
  III}}(1996)}]{Kong_Brooks:1996}%
  \BibitemOpen
  \bibfield  {author} {\bibinfo {author} {\bibfnamefont {X.}~\bibnamefont
  {Kong}}\ and\ \bibinfo {author} {\bibfnamefont {C.~L.}\ \bibnamefont {{Brooks
  III}}},\ }\href {http://dx.doi.org/10.1063/1.472109} {\bibfield  {journal}
  {\bibinfo  {journal} {J. Chem. Phys.}\ }\textbf {\bibinfo {volume} {105}},\
  \bibinfo {eid} {2414} (\bibinfo {year} {1996})}\BibitemShut {NoStop}%
\bibitem [{\citenamefont {Guo}, \citenamefont {{Brooks III}},\ and\
  \citenamefont {Kong}(1998)}]{Guo_Kong:1998}%
  \BibitemOpen
  \bibfield  {author} {\bibinfo {author} {\bibfnamefont {Z.}~\bibnamefont
  {Guo}}, \bibinfo {author} {\bibfnamefont {C.~L.}\ \bibnamefont {{Brooks
  III}}}, \ and\ \bibinfo {author} {\bibfnamefont {X.}~\bibnamefont {Kong}},\
  }\href {http://pubs.acs.org/doi/abs/10.1021/jp972699%2B} {\bibfield
  {journal} {\bibinfo  {journal} {J. Phys. Chem. B}\ }\textbf {\bibinfo
  {volume} {102}},\ \bibinfo {eid} {2032} (\bibinfo {year} {1998})}\BibitemShut
  {NoStop}%
\bibitem [{\citenamefont {Bitetti-Putzer}, \citenamefont {Yang},\ and\
  \citenamefont {Karplus}(2003)}]{Bitetti-Putzer_Karplus:2003}%
  \BibitemOpen
  \bibfield  {author} {\bibinfo {author} {\bibfnamefont {R.}~\bibnamefont
  {Bitetti-Putzer}}, \bibinfo {author} {\bibfnamefont {W.}~\bibnamefont
  {Yang}}, \ and\ \bibinfo {author} {\bibfnamefont {M.}~\bibnamefont
  {Karplus}},\ }\href@noop {} {\bibfield  {journal} {\bibinfo  {journal} {Chem.
  Phys. Lett.}\ }\textbf {\bibinfo {volume} {377}},\ \bibinfo {eid} {633}
  (\bibinfo {year} {2003})}\BibitemShut {NoStop}%
\bibitem [{\citenamefont {Osawa}\ \emph {et~al.}(2010)\citenamefont {Osawa},
  \citenamefont {Futakuchi}, \citenamefont {Imahori},\ and\ \citenamefont
  {Lee}}]{Osawa_Lee:2010}%
  \BibitemOpen
  \bibfield  {author} {\bibinfo {author} {\bibfnamefont {T.}~\bibnamefont
  {Osawa}}, \bibinfo {author} {\bibfnamefont {T.}~\bibnamefont {Futakuchi}},
  \bibinfo {author} {\bibfnamefont {T.}~\bibnamefont {Imahori}}, \ and\
  \bibinfo {author} {\bibfnamefont {I.-Y.~S.}\ \bibnamefont {Lee}},\
  }\href@noop {} {\bibfield  {journal} {\bibinfo  {journal} {J. Mol. Catal. A}\
  }\textbf {\bibinfo {volume} {320}},\ \bibinfo {eid} {68} (\bibinfo {year}
  {2010})}\BibitemShut {NoStop}%
\bibitem [{\citenamefont {Hargreaves}\ \emph {et~al.}(2002)\citenamefont
  {Hargreaves}, \citenamefont {Hutchings}, \citenamefont {Joyner},\ and\
  \citenamefont {Taylor}}]{Hargreaves_Taylor:2002}%
  \BibitemOpen
  \bibfield  {author} {\bibinfo {author} {\bibfnamefont {J.~S.~J.}\
  \bibnamefont {Hargreaves}}, \bibinfo {author} {\bibfnamefont {G.~J.}\
  \bibnamefont {Hutchings}}, \bibinfo {author} {\bibfnamefont {R.~W.}\
  \bibnamefont {Joyner}}, \ and\ \bibinfo {author} {\bibfnamefont {S.~H.}\
  \bibnamefont {Taylor}},\ }\href
  {http://www.sciencedirect.com/science/article/pii/S0926860X01009358}
  {\bibfield  {journal} {\bibinfo  {journal} {Appl. Catal. A}\ }\textbf
  {\bibinfo {volume} {227}},\ \bibinfo {eid} {191} (\bibinfo {year}
  {2002})}\BibitemShut {NoStop}%
\bibitem [{\citenamefont {Lynch}, \citenamefont {Mielke},\ and\ \citenamefont
  {Truhlar}(2005)}]{Lynch_Truhlar:2005}%
  \BibitemOpen
  \bibfield  {author} {\bibinfo {author} {\bibfnamefont {V.~A.}\ \bibnamefont
  {Lynch}}, \bibinfo {author} {\bibfnamefont {S.~L.}\ \bibnamefont {Mielke}}, \
  and\ \bibinfo {author} {\bibfnamefont {D.~G.}\ \bibnamefont {Truhlar}},\
  }\href {http://pubs.acs.org/doi/pdf/10.1021/jp051742n} {\bibfield  {journal}
  {\bibinfo  {journal} {J. Phys. Chem}\ }\textbf {\bibinfo {volume} {109}},\
  \bibinfo {eid} {10092} (\bibinfo {year} {2005})}\BibitemShut {NoStop}%
\bibitem [{\citenamefont {Predescu}\ and\ \citenamefont
  {Doll}(2002)}]{Predescu_Doll:2002}%
  \BibitemOpen
  \bibfield  {author} {\bibinfo {author} {\bibfnamefont {C.}~\bibnamefont
  {Predescu}}\ and\ \bibinfo {author} {\bibfnamefont {J.~D.}\ \bibnamefont
  {Doll}},\ }\href {http://dx.doi.org/10.1063/1.1509058} {\bibfield  {journal}
  {\bibinfo  {journal} {J. Chem. Phys.}\ }\textbf {\bibinfo {volume} {117}},\
  \bibinfo {pages} {7448} (\bibinfo {year} {2002})}\BibitemShut {NoStop}%
\bibitem [{\citenamefont {Yamamoto}(2005)}]{Yamamoto:2005}%
  \BibitemOpen
  \bibfield  {author} {\bibinfo {author} {\bibfnamefont {T.}~\bibnamefont
  {Yamamoto}},\ }\href {http://dx.doi.org/10.1063/1.2013257} {\bibfield
  {journal} {\bibinfo  {journal} {J. Chem. Phys.}\ }\textbf {\bibinfo {volume}
  {123}},\ \bibinfo {eid} {104101} (\bibinfo {year} {2005})}\BibitemShut
  {NoStop}%
\bibitem [{\citenamefont {K\"{a}stner}\ and\ \citenamefont
  {Thiel}(2005)}]{Kastner_Thiel:2005}%
  \BibitemOpen
  \bibfield  {author} {\bibinfo {author} {\bibfnamefont {J.}~\bibnamefont
  {K\"{a}stner}}\ and\ \bibinfo {author} {\bibfnamefont {W.}~\bibnamefont
  {Thiel}},\ }\href {http://dx.doi.org/10.1063/1.2052648} {\bibfield  {journal}
  {\bibinfo  {journal} {J. Chem. Phys.}\ }\textbf {\bibinfo {volume} {123}},\
  \bibinfo {eid} {144104} (\bibinfo {year} {2005})}\BibitemShut {NoStop}%
\bibitem [{\citenamefont {K\"{a}stner}(2009)}]{Kastner:2009}%
  \BibitemOpen
  \bibfield  {author} {\bibinfo {author} {\bibfnamefont {J.}~\bibnamefont
  {K\"{a}stner}},\ }\href {http://dx.doi.org/10.1063/1.3175798} {\bibfield
  {journal} {\bibinfo  {journal} {J. Chem. Phys.}\ }\textbf {\bibinfo {volume}
  {131}},\ \bibinfo {eid} {034109} (\bibinfo {year} {2009})}\BibitemShut
  {NoStop}%
\bibitem [{\citenamefont {K\"{a}stner}(2012)}]{Kastner:2012}%
  \BibitemOpen
  \bibfield  {author} {\bibinfo {author} {\bibfnamefont {J.}~\bibnamefont
  {K\"{a}stner}},\ }\href {http://dx.doi.org/10.1063/1.4729373} {\bibfield
  {journal} {\bibinfo  {journal} {J. Chem. Phys.}\ }\textbf {\bibinfo {volume}
  {136}},\ \bibinfo {eid} {234102} (\bibinfo {year} {2012})}\BibitemShut
  {NoStop}%
\bibitem [{\citenamefont {Darve}\ and\ \citenamefont
  {Pohorille}(2001)}]{Darve_Pohorille:2001}%
  \BibitemOpen
  \bibfield  {author} {\bibinfo {author} {\bibfnamefont {E.}~\bibnamefont
  {Darve}}\ and\ \bibinfo {author} {\bibfnamefont {A.}~\bibnamefont
  {Pohorille}},\ }\href@noop {} {\bibfield  {journal} {\bibinfo  {journal} {J.
  Chem. Phys.}\ }\textbf {\bibinfo {volume} {115}},\ \bibinfo {eid} {9169}
  (\bibinfo {year} {2001})}\BibitemShut {NoStop}%
\bibitem [{\citenamefont {Darve}, \citenamefont {Rodr{\'i}gues-G{\'o}mez},\
  and\ \citenamefont {Pohorille}(2008)}]{Darve_Pohorille:2008}%
  \BibitemOpen
  \bibfield  {author} {\bibinfo {author} {\bibfnamefont {E.}~\bibnamefont
  {Darve}}, \bibinfo {author} {\bibfnamefont {D.}~\bibnamefont
  {Rodr{\'i}gues-G{\'o}mez}}, \ and\ \bibinfo {author} {\bibfnamefont
  {A.}~\bibnamefont {Pohorille}},\ }\href@noop {} {\bibfield  {journal}
  {\bibinfo  {journal} {J. Chem. Phys.}\ }\textbf {\bibinfo {volume} {128}},\
  \bibinfo {eid} {144120} (\bibinfo {year} {2008})}\BibitemShut {NoStop}%
\bibitem [{\citenamefont {Comer}\ \emph {et~al.}(2015)\citenamefont {Comer},
  \citenamefont {Gumbart}, \citenamefont {H\'{e}nin}, \citenamefont
  {Leli{\`e}vre}, \citenamefont {Pohorille},\ and\ \citenamefont
  {Chipot}}]{Comer_Chipot:2015}%
  \BibitemOpen
  \bibfield  {author} {\bibinfo {author} {\bibfnamefont {J.}~\bibnamefont
  {Comer}}, \bibinfo {author} {\bibfnamefont {J.~C.}\ \bibnamefont {Gumbart}},
  \bibinfo {author} {\bibfnamefont {J.}~\bibnamefont {H\'{e}nin}}, \bibinfo
  {author} {\bibfnamefont {T.}~\bibnamefont {Leli{\`e}vre}}, \bibinfo {author}
  {\bibfnamefont {A.}~\bibnamefont {Pohorille}}, \ and\ \bibinfo {author}
  {\bibfnamefont {C.}~\bibnamefont {Chipot}},\ }\href@noop {} {\bibfield
  {journal} {\bibinfo  {journal} {J. Phys. Chem. B}\ }\textbf {\bibinfo
  {volume} {119}},\ \bibinfo {eid} {1129} (\bibinfo {year} {2015})}\BibitemShut
  {NoStop}%
\bibitem [{\citenamefont {Mezei}(1987)}]{Mezei:1987}%
  \BibitemOpen
  \bibfield  {author} {\bibinfo {author} {\bibfnamefont {M.}~\bibnamefont
  {Mezei}},\ }\href
  {http://www.sciencedirect.com/science/article/pii/0021999187900544}
  {\bibfield  {journal} {\bibinfo  {journal} {J. Comput. Phys.}\ }\textbf
  {\bibinfo {volume} {68}},\ \bibinfo {eid} {237} (\bibinfo {year}
  {1987})}\BibitemShut {NoStop}%
\bibitem [{\citenamefont {Hooft}, \citenamefont {van Eijck},\ and\
  \citenamefont {Kroon}(1992)}]{Hooft_Kroon:1992}%
  \BibitemOpen
  \bibfield  {author} {\bibinfo {author} {\bibfnamefont {R.~W.~W.}\
  \bibnamefont {Hooft}}, \bibinfo {author} {\bibfnamefont {B.~P.}\ \bibnamefont
  {van Eijck}}, \ and\ \bibinfo {author} {\bibfnamefont {J.}~\bibnamefont
  {Kroon}},\ }\href {http://dx.doi.org/10.1063/1.463947} {\bibfield  {journal}
  {\bibinfo  {journal} {J. Chem. Phys.}\ }\textbf {\bibinfo {volume} {97}},\
  \bibinfo {eid} {6690} (\bibinfo {year} {1992})}\BibitemShut {NoStop}%
\bibitem [{\citenamefont {Bartels}\ and\ \citenamefont
  {Karplus}(1997)}]{Bartels_Karplus:1997}%
  \BibitemOpen
  \bibfield  {author} {\bibinfo {author} {\bibfnamefont {C.}~\bibnamefont
  {Bartels}}\ and\ \bibinfo {author} {\bibfnamefont {M.}~\bibnamefont
  {Karplus}},\ }\href
  {http://onlinelibrary.wiley.com/doi/10.1002/%28SICI%291096-987X%28199709%2918:12%3C1450::AID-JCC3%3E3.0.CO;2-I/full}
  {\bibfield  {journal} {\bibinfo  {journal} {J. Comput. Chem.}\ }\textbf
  {\bibinfo {volume} {18}},\ \bibinfo {eid} {1450} (\bibinfo {year}
  {1997})}\BibitemShut {NoStop}%
\bibitem [{\citenamefont {Micheletti}, \citenamefont {Laio},\ and\
  \citenamefont {Parrinello}(2004)}]{Micheletti_Parrinello:2004}%
  \BibitemOpen
  \bibfield  {author} {\bibinfo {author} {\bibfnamefont {C.}~\bibnamefont
  {Micheletti}}, \bibinfo {author} {\bibfnamefont {A.}~\bibnamefont {Laio}}, \
  and\ \bibinfo {author} {\bibfnamefont {M.}~\bibnamefont {Parrinello}},\
  }\href@noop {} {\bibfield  {journal} {\bibinfo  {journal} {Phys. Rev. Lett.}\
  }\textbf {\bibinfo {volume} {92}},\ \bibinfo {eid} {170601} (\bibinfo {year}
  {2004})}\BibitemShut {NoStop}%
\bibitem [{\citenamefont {Laio}\ \emph {et~al.}(2005)\citenamefont {Laio},
  \citenamefont {Rodriguez-Fortea}, \citenamefont {Gervasio}, \citenamefont
  {Ceccarelli},\ and\ \citenamefont {Parrinello}}]{Laio_Parrinello:2005}%
  \BibitemOpen
  \bibfield  {author} {\bibinfo {author} {\bibfnamefont {A.}~\bibnamefont
  {Laio}}, \bibinfo {author} {\bibfnamefont {A.}~\bibnamefont
  {Rodriguez-Fortea}}, \bibinfo {author} {\bibfnamefont {F.~L.}\ \bibnamefont
  {Gervasio}}, \bibinfo {author} {\bibfnamefont {M.}~\bibnamefont
  {Ceccarelli}}, \ and\ \bibinfo {author} {\bibfnamefont {M.}~\bibnamefont
  {Parrinello}},\ }\href@noop {} {\bibfield  {journal} {\bibinfo  {journal} {J.
  Phys. Chem. B}\ }\textbf {\bibinfo {volume} {109}},\ \bibinfo {eid} {6714}
  (\bibinfo {year} {2005})}\BibitemShut {NoStop}%
\bibitem [{\citenamefont {Schweizer}\ \emph {et~al.}(1981)\citenamefont
  {Schweizer}, \citenamefont {Stratt}, \citenamefont {Chandler},\ and\
  \citenamefont {Wolynes}}]{Schweizer_Wolynes:1981}%
  \BibitemOpen
  \bibfield  {author} {\bibinfo {author} {\bibfnamefont {K.~S.}\ \bibnamefont
  {Schweizer}}, \bibinfo {author} {\bibfnamefont {R.~M.}\ \bibnamefont
  {Stratt}}, \bibinfo {author} {\bibfnamefont {D.}~\bibnamefont {Chandler}}, \
  and\ \bibinfo {author} {\bibfnamefont {P.~G.}\ \bibnamefont {Wolynes}},\
  }\href
  {http://scitation.aip.org/content/aip/journal/jcp/75/3/10.1063/1.442141}
  {\bibfield  {journal} {\bibinfo  {journal} {J. Chem. Phys.}\ }\textbf
  {\bibinfo {volume} {75}},\ \bibinfo {eid} {1347} (\bibinfo {year}
  {1981})}\BibitemShut {NoStop}%
\bibitem [{\citenamefont {Flyvbjerg}\ and\ \citenamefont
  {Petersen}(1989)}]{Flyvbjerg_Petersen:1989}%
  \BibitemOpen
  \bibfield  {author} {\bibinfo {author} {\bibfnamefont {H.}~\bibnamefont
  {Flyvbjerg}}\ and\ \bibinfo {author} {\bibfnamefont {H.~G.}\ \bibnamefont
  {Petersen}},\ }\href {http://dx.doi.org/10.1063/1.457480} {\bibfield
  {journal} {\bibinfo  {journal} {J. Chem. Phys.}\ }\textbf {\bibinfo {volume}
  {91}},\ \bibinfo {eid} {461} (\bibinfo {year} {1989})}\BibitemShut {NoStop}%
\bibitem [{\citenamefont {Herman}, \citenamefont {Bruskin},\ and\ \citenamefont
  {Berne}(1982)}]{Herman_Berne:1982}%
  \BibitemOpen
  \bibfield  {author} {\bibinfo {author} {\bibfnamefont {M.~F.}\ \bibnamefont
  {Herman}}, \bibinfo {author} {\bibfnamefont {E.~J.}\ \bibnamefont {Bruskin}},
  \ and\ \bibinfo {author} {\bibfnamefont {B.~J.}\ \bibnamefont {Berne}},\
  }\href
  {http://scitation.aip.org/content/aip/journal/jcp/76/10/10.1063/1.442815}
  {\bibfield  {journal} {\bibinfo  {journal} {J. Chem. Phys.}\ }\textbf
  {\bibinfo {volume} {76}},\ \bibinfo {eid} {5150} (\bibinfo {year}
  {1982})}\BibitemShut {NoStop}%
\bibitem [{\citenamefont {Cao}\ and\ \citenamefont
  {Berne}(1993)}]{Cao_Berne:1993}%
  \BibitemOpen
  \bibfield  {author} {\bibinfo {author} {\bibfnamefont {J.}~\bibnamefont
  {Cao}}\ and\ \bibinfo {author} {\bibfnamefont {B.~J.}\ \bibnamefont
  {Berne}},\ }\href
  {http://scitation.aip.org/content/aip/journal/jcp/99/4/10.1063/1.465198}
  {\bibfield  {journal} {\bibinfo  {journal} {J. Chem. Phys.}\ }\textbf
  {\bibinfo {volume} {99}},\ \bibinfo {eid} {2902} (\bibinfo {year}
  {1993})}\BibitemShut {NoStop}%
\bibitem [{\citenamefont {Schwenke}\ and\ \citenamefont
  {Partridge}(2001)}]{Schwenke_Partridge:2001}%
  \BibitemOpen
  \bibfield  {author} {\bibinfo {author} {\bibfnamefont {D.~W.}\ \bibnamefont
  {Schwenke}}\ and\ \bibinfo {author} {\bibfnamefont {H.}~\bibnamefont
  {Partridge}},\ }\href@noop {} {\bibfield  {journal} {\bibinfo  {journal}
  {Spectrochim. Acta A}\ }\textbf {\bibinfo {volume} {57}},\ \bibinfo {eid}
  {887} (\bibinfo {year} {2001})}\BibitemShut {NoStop}%
\bibitem [{\citenamefont {Duchovic}\ \emph {et~al.}()\citenamefont {Duchovic},
  \citenamefont {Volobuev}, \citenamefont {Lynch}, \citenamefont {Jasper},
  \citenamefont {Truhlar}, \citenamefont {Allison}, \citenamefont {Wagner},
  \citenamefont {Garrett}, \citenamefont {Espinosa-Garc\'{i}a},\ and\
  \citenamefont {Corchado}}]{POTLIB}%
  \BibitemOpen
  \bibfield  {author} {\bibinfo {author} {\bibfnamefont {R.~J.}\ \bibnamefont
  {Duchovic}}, \bibinfo {author} {\bibfnamefont {Y.~L.}\ \bibnamefont
  {Volobuev}}, \bibinfo {author} {\bibfnamefont {G.~C.}\ \bibnamefont {Lynch}},
  \bibinfo {author} {\bibfnamefont {A.~W.}\ \bibnamefont {Jasper}}, \bibinfo
  {author} {\bibfnamefont {D.~G.}\ \bibnamefont {Truhlar}}, \bibinfo {author}
  {\bibfnamefont {T.~C.}\ \bibnamefont {Allison}}, \bibinfo {author}
  {\bibfnamefont {A.~F.}\ \bibnamefont {Wagner}}, \bibinfo {author}
  {\bibfnamefont {B.~C.}\ \bibnamefont {Garrett}}, \bibinfo {author}
  {\bibfnamefont {J.}~\bibnamefont {Espinosa-Garc\'{i}a}}, \ and\ \bibinfo
  {author} {\bibfnamefont {J.~C.}\ \bibnamefont {Corchado}},\ }\href@noop {}
  {\emph {\bibinfo {title} {\textup{POTLIB,
  http://comp.chem.umn.edu/potlib}}}}\BibitemShut {NoStop}%
\bibitem [{\citenamefont {Sprik}, \citenamefont {Klein},\ and\ \citenamefont
  {Chandler}(1985{\natexlab{a}})}]{Sprik_Chandler:1985}%
  \BibitemOpen
  \bibfield  {author} {\bibinfo {author} {\bibfnamefont {M.}~\bibnamefont
  {Sprik}}, \bibinfo {author} {\bibfnamefont {M.~L.}\ \bibnamefont {Klein}}, \
  and\ \bibinfo {author} {\bibfnamefont {D.}~\bibnamefont {Chandler}},\
  }\href@noop {} {\bibfield  {journal} {\bibinfo  {journal} {Phys. Rev. B}\
  }\textbf {\bibinfo {volume} {31}},\ \bibinfo {eid} {4234} (\bibinfo {year}
  {1985}{\natexlab{a}})}\BibitemShut {NoStop}%
\bibitem [{\citenamefont {Sprik}, \citenamefont {Klein},\ and\ \citenamefont
  {Chandler}(1985{\natexlab{b}})}]{Sprik_Chandler:1985_1}%
  \BibitemOpen
  \bibfield  {author} {\bibinfo {author} {\bibfnamefont {M.}~\bibnamefont
  {Sprik}}, \bibinfo {author} {\bibfnamefont {M.~L.}\ \bibnamefont {Klein}}, \
  and\ \bibinfo {author} {\bibfnamefont {D.}~\bibnamefont {Chandler}},\
  }\href@noop {} {\bibfield  {journal} {\bibinfo  {journal} {Phys. Rev. B}\
  }\textbf {\bibinfo {volume} {32}},\ \bibinfo {eid} {545} (\bibinfo {year}
  {1985}{\natexlab{b}})}\BibitemShut {NoStop}%
\bibitem [{\citenamefont {Abrams}, \citenamefont {Rosso},\ and\ \citenamefont
  {Tuckerman}(2006)}]{Abrams_Tuckerman:2006}%
  \BibitemOpen
  \bibfield  {author} {\bibinfo {author} {\bibfnamefont {J.~B.}\ \bibnamefont
  {Abrams}}, \bibinfo {author} {\bibfnamefont {L.}~\bibnamefont {Rosso}}, \
  and\ \bibinfo {author} {\bibfnamefont {M.~E.}\ \bibnamefont {Tuckerman}},\
  }\href@noop {} {\bibfield  {journal} {\bibinfo  {journal} {J. Chem. Phys.}\
  }\textbf {\bibinfo {volume} {125}},\ \bibinfo {eid} {074115} (\bibinfo {year}
  {2006})}\BibitemShut {NoStop}%
\bibitem [{\citenamefont {Wu}, \citenamefont {Hu},\ and\ \citenamefont
  {Yang}(2011)}]{Wu_Yang:2011}%
  \BibitemOpen
  \bibfield  {author} {\bibinfo {author} {\bibfnamefont {P.}~\bibnamefont
  {Wu}}, \bibinfo {author} {\bibfnamefont {X.}~\bibnamefont {Hu}}, \ and\
  \bibinfo {author} {\bibfnamefont {W.}~\bibnamefont {Yang}},\ }\href@noop {}
  {\bibfield  {journal} {\bibinfo  {journal} {J. Phys. Chem. Lett.}\ }\textbf
  {\bibinfo {volume} {2}},\ \bibinfo {eid} {2099} (\bibinfo {year}
  {2011})}\BibitemShut {NoStop}%
\bibitem [{\citenamefont {Cheng}\ and\ \citenamefont
  {Ceriotti}(2015)}]{Cheng_Ceriotti:2015}%
  \BibitemOpen
  \bibfield  {author} {\bibinfo {author} {\bibfnamefont {B.}~\bibnamefont
  {Cheng}}\ and\ \bibinfo {author} {\bibfnamefont {M.}~\bibnamefont
  {Ceriotti}},\ }\href {http://dx.doi.org/10.1063/1.4904293} {\bibfield
  {journal} {\bibinfo  {journal} {J. Chem. Phys.}\ }\textbf {\bibinfo {volume}
  {141}},\ \bibinfo {eid} {244112} (\bibinfo {year} {2015})}\BibitemShut
  {NoStop}%
\bibitem [{\citenamefont {Karandashev}\ and\ \citenamefont
  {Van\'{i}\v{c}ek}()}]{Karandashev_Vanicek:2017b}%
  \BibitemOpen
  \bibfield  {author} {\bibinfo {author} {\bibfnamefont {K.}~\bibnamefont
  {Karandashev}}\ and\ \bibinfo {author} {\bibfnamefont {J.}~\bibnamefont
  {Van\'{i}\v{c}ek}},\ }\href@noop {} {\enquote {\bibinfo {title} {Accelerating
  equilibrium isotope effect calculations: {II}. stochastic implementation of
  direct estimators},}\ }\bibinfo {note} {{J. Chem. Phys.}, {i}n
  preparation}\BibitemShut {NoStop}%
\bibitem [{\citenamefont {Takahashi}\ and\ \citenamefont
  {Imada}(1984)}]{Takahashi_Imada:1984}%
  \BibitemOpen
  \bibfield  {author} {\bibinfo {author} {\bibfnamefont {M.}~\bibnamefont
  {Takahashi}}\ and\ \bibinfo {author} {\bibfnamefont {M.}~\bibnamefont
  {Imada}},\ }\href {http://dx.doi.org/10.1143/JPSJ.53.963} {\bibfield
  {journal} {\bibinfo  {journal} {J. Phys. Soc. Jpn.}\ }\textbf {\bibinfo
  {volume} {53}},\ \bibinfo {eid} {963} (\bibinfo {year} {1984})}\BibitemShut
  {NoStop}%
\bibitem [{\citenamefont {Suzuki}(1995)}]{Suzuki:1995}%
  \BibitemOpen
  \bibfield  {author} {\bibinfo {author} {\bibfnamefont {M.}~\bibnamefont
  {Suzuki}},\ }\href {http://dx.doi.org/10.1016/0375-9601(95)00266-6}
  {\bibfield  {journal} {\bibinfo  {journal} {Physics Letters A}\ }\textbf
  {\bibinfo {volume} {201}},\ \bibinfo {eid} {425} (\bibinfo {year}
  {1995})}\BibitemShut {NoStop}%
\bibitem [{\citenamefont {Chin}(1997)}]{Chin:1997}%
  \BibitemOpen
  \bibfield  {author} {\bibinfo {author} {\bibfnamefont {S.~A.}\ \bibnamefont
  {Chin}},\ }\href
  {http://www.sciencedirect.com/science/article/pii/S0375960197000030}
  {\bibfield  {journal} {\bibinfo  {journal} {Phys. Lett. A}\ }\textbf
  {\bibinfo {volume} {226}},\ \bibinfo {eid} {344} (\bibinfo {year}
  {1997})}\BibitemShut {NoStop}%
\bibitem [{\citenamefont {Jang}, \citenamefont {Jang},\ and\ \citenamefont
  {Voth}(2001)}]{Jang_Voth:2001}%
  \BibitemOpen
  \bibfield  {author} {\bibinfo {author} {\bibfnamefont {S.}~\bibnamefont
  {Jang}}, \bibinfo {author} {\bibfnamefont {S.}~\bibnamefont {Jang}}, \ and\
  \bibinfo {author} {\bibfnamefont {G.~A.}\ \bibnamefont {Voth}},\ }\href
  {http://dx.doi.org/10.1063/1.1410117} {\bibfield  {journal} {\bibinfo
  {journal} {J. Chem. Phys.}\ }\textbf {\bibinfo {volume} {115}},\ \bibinfo
  {eid} {7832} (\bibinfo {year} {2001})}\BibitemShut {NoStop}%
\bibitem [{\citenamefont {P\'{e}rez}\ and\ \citenamefont
  {Tuckerman}(2011)}]{Perez_Tuckerman:2011}%
  \BibitemOpen
  \bibfield  {author} {\bibinfo {author} {\bibfnamefont {A.}~\bibnamefont
  {P\'{e}rez}}\ and\ \bibinfo {author} {\bibfnamefont {M.~E.}\ \bibnamefont
  {Tuckerman}},\ }\href {http://dx.doi.org/10.1063/1.3609120} {\bibfield
  {journal} {\bibinfo  {journal} {J. Chem. Phys.}\ }\textbf {\bibinfo {volume}
  {135}},\ \bibinfo {eid} {064104} (\bibinfo {year} {2011})}\BibitemShut
  {NoStop}%
\bibitem [{\citenamefont {Buchowiecki}\ and\ \citenamefont
  {Van\'{i}\v{c}ek}(2013)}]{Buchowiecki_Vanicek:2013}%
  \BibitemOpen
  \bibfield  {author} {\bibinfo {author} {\bibfnamefont {M.}~\bibnamefont
  {Buchowiecki}}\ and\ \bibinfo {author} {\bibfnamefont {J.}~\bibnamefont
  {Van\'{i}\v{c}ek}},\ }\href
  {http://www.sciencedirect.com/science/article/pii/S0009261413012621}
  {\bibfield  {journal} {\bibinfo  {journal} {Chem. Phys. Lett.}\ }\textbf
  {\bibinfo {volume} {588}},\ \bibinfo {eid} {11} (\bibinfo {year}
  {2013})}\BibitemShut {NoStop}%
\bibitem [{\citenamefont {Karandashev}\ and\ \citenamefont
  {Van\'{i}\v{c}ek}(2015)}]{Karandashev_Vanicek:2015}%
  \BibitemOpen
  \bibfield  {author} {\bibinfo {author} {\bibfnamefont {K.}~\bibnamefont
  {Karandashev}}\ and\ \bibinfo {author} {\bibfnamefont {J.}~\bibnamefont
  {Van\'{i}\v{c}ek}},\ }\href {http://dx.doi.org/10.1063/1.4935701} {\bibfield
  {journal} {\bibinfo  {journal} {J. Chem. Phys.}\ }\textbf {\bibinfo {volume}
  {143}},\ \bibinfo {eid} {194104} (\bibinfo {year} {2015})}\BibitemShut
  {NoStop}%
\bibitem [{\citenamefont {Coalson}, \citenamefont {Freeman},\ and\
  \citenamefont {Doll}(1986)}]{Coalson_Doll:1986}%
  \BibitemOpen
  \bibfield  {author} {\bibinfo {author} {\bibfnamefont {R.~D.}\ \bibnamefont
  {Coalson}}, \bibinfo {author} {\bibfnamefont {D.~L.}\ \bibnamefont
  {Freeman}}, \ and\ \bibinfo {author} {\bibfnamefont {J.~D.}\ \bibnamefont
  {Doll}},\ }\href@noop {} {\bibfield  {journal} {\bibinfo  {journal} {J. Chem.
  Phys.}\ }\textbf {\bibinfo {volume} {85}},\ \bibinfo {eid} {4567} (\bibinfo
  {year} {1986})}\BibitemShut {NoStop}%
\bibitem [{\citenamefont {Ersoy}(1985)}]{Ersoy:1985}%
  \BibitemOpen
  \bibfield  {author} {\bibinfo {author} {\bibfnamefont {O.}~\bibnamefont
  {Ersoy}},\ }\href@noop {} {\bibfield  {journal} {\bibinfo  {journal} {{IEEE}
  Trans. Acoust., Speech, Signal Process.}\ }\textbf {\bibinfo {volume} {33}},\
  \bibinfo {eid} {880} (\bibinfo {year} {1985})}\BibitemShut {NoStop}%
\end{thebibliography}

\end{document}